\documentclass[a4paper]{IEEEtran}
\usepackage{subfigure}
\usepackage{graphicx}
\usepackage{amsmath}
\usepackage{epsfig}
\usepackage{amsfonts}
\usepackage{amssymb}
\usepackage{gensymb}
\usepackage{epstopdf}
\usepackage[top=2.8cm, left=1.5cm, right=1.5cm, bottom=2.2cm]{geometry}

\newcommand{\argmax}{\mathop{\text{argmax}}}
\newcommand{\argmin}{\mathop{\text{argmin}}}

\newcommand{\vh}{{\bf h}}
\newcommand{\vv}{{\bf v}}

\newcommand{\vx}{{\bf x}}
\newcommand{\vy}{{\bf y}}
\newcommand{\vs}{{\bf s}}
\newcommand{\vn}{{\bf n}}

\newcommand{\mh}{{\bf H}}
\newcommand{\mf}{{\bf F}}
\newcommand{\my}{{\bf Y}}
\newcommand{\mx}{{\bf X}}

\newcommand{\se}{{\mathbb E}}

\newcommand{\Define}{\triangleq}

\begin{document}
\title{{\LARGE
Quad-LED and Dual-LED Complex Modulation for Visible Light 
Communication }}
\author{T. Lakshmi Narasimhan$^{\dagger}$, R. Tejaswi, and A. Chockalingam \\
$\dagger$ Presently with Department of EECS, Syracuse University, Syracuse,
NY 13244, USA \\
Department of ECE, Indian Institute of Science, Bangalore 560012 }
\maketitle

\thispagestyle{empty} 
\begin{abstract}
In this paper, we propose simple and novel complex modulation techniques 
that exploit the spatial domain to transmit complex-valued modulation 
symbols in visible light wireless communication. The idea is to use 
multiple light emitting diodes (LEDs) to convey the real and imaginary 
parts of a complex modulation symbol and their sign information, or, 
alternately, to convey the magnitude and phase of a complex symbol. The 
proposed techniques are termed as {\em quad-LED complex modulation (QCM)} 
and {\em dual-LED complex modulation (DCM)}. The proposed QCM scheme uses 
four LEDs (hence the name `quad-LED'); while the magnitudes of the real 
and imaginary parts are conveyed through intensity modulation of LEDs, 
the sign information is conveyed through spatial indexing of LEDs.
The proposed DCM scheme, on the other hand, exploits the polar 
representation of a complex symbol; it uses only two LEDs (hence the 
name `dual-LED'), one LED to map the magnitude and another LED to map 
the phase of a complex modulation symbol. These techniques do not need 
Hermitian symmetry operation to generate LED compatible positive real 
transmit signals. We present zero-forcing and minimum distance detectors 
and their performance for QCM-OFDM and DCM-OFDM. We further propose 
another modulation scheme, termed as SM-DCM {\em (spatial modulation-DCM)} 
scheme, which brings in the advantage of spatial modulation (SM) to DCM. 
The proposed SM-DCM scheme uses two DCM BLOCKs with two LEDs in each BLOCK, 
and an index bit decides which among the two BLOCKs will be used in a given 
channel use. We study the bit error rate (BER) performance of the proposed 
schemes through analysis and simulations. Using tight analytical BER upper 
bounds and spatial distribution of the received signal-to-noise ratios, we 
compute and plot the achievable rate contours for a given target BER in QCM, 
DCM, and SM-DCM.

\end{abstract}

\pagestyle{empty} 
\section{Introduction}
\label{sec1}
Wireless communication using visible light wavelengths (400 to 700 nm) 
in indoor local area network environments is emerging as a promising 
area of research \cite{haas1}. Visible light communication (VLC) is 
evolving as an appealing complementary technology to radio frequency (RF) 
communication technology \cite{brien}. In VLC, simple and inexpensive 
light emitting diodes (LED) and photo diodes (PD) act as signal 
transmitters and receptors, respectively, replacing more complex and 
expensive transmit/receive RF hardware and antennas in RF wireless
communication systems. Other favorable features in VLC include 
availability of abundant visible light spectrum at no cost, no 
licensing/RF radiation issues, and inherent security in closed-room 
applications. The possibility of using the same LEDs to simultaneously 
provide both energy-efficient lighting as well as high-speed short-range
communication is another attractive feature.

The potential to use multiple LEDs and PDs in multiple-input 
multiple-output (MIMO) array configurations has enthused MIMO wireless 
researchers to take special interest in VLC \cite{jean}-\cite{gsm_vlc}. 
Signaling schemes considered in multiple-LED VLC include 
space shift keying (SSK) and its generalization (GSSK), where the 
ON/OFF status of the LEDs and the indices of the LEDs which are ON 
convey information bits \cite{vlc4},\cite{vlc3a}. Other multiple-LED 
signaling schemes considered in the literature include spatial 
multiplexing (SMP), spatial modulation (SM), and generalized spatial 
modulation (GSM) \cite{vlc5},\cite{vlc1},\cite{gsm_vlc},\cite{smvlc}. 
These works have considered real signal sets like $M$-ary pulse amplitude 
modulation (PAM) with positive-valued signal points in line with the 
need for the transmit signal in VLC to be positive and real-valued to 
intensity modulate the LEDs.  

The VLC channel between an LED and a photo detector in indoor 
environments can be a multipath channel. The multipath effects 
can be mitigated by using orthogonal frequency division 
multiplexing (OFDM). The use of complex signal sets like $M$-ary 
quadrature amplitude modulation (QAM) along with OFDM in VLC is 
studied extensively in the literature \cite{ofdm1}-\cite{indc}. 
Techniques reported in these works include DC-biased optical (DCO) 
OFDM \cite{dco2}, asymmetrically clipped optical (ACO) OFDM 
\cite{aco1}-\cite{aco2}, flip OFDM \cite{flip1},\cite{flip2}, 
non-DC biased (NDC) OFDM \cite{ndc}, and index modulation for 
NDC OFDM \cite{indc}. A key constraint in the above techniques, 
however, is that they perform Hermitian symmetry operation on the 
QAM symbol vector at the IFFT input so that the IFFT output would 
be positive and real-valued. A consequence of this is that $N$ 
channel uses are needed to send $N/2$ symbols. 

In this paper, we propose two simple and novel complex modulation 
techniques for VLC using multiple LEDs, which do not need Hermitian 
symmetry operation. The proposed schemes exploit the spatial dimension 
to convey complex-valued modulation symbols. 
\begin{itemize}
\item The first proposed idea 
is to use four LEDs to form a single modulation unit that simultaneously 
conveys the real and imaginary parts of a complex modulation symbol and 
their sign information. While the magnitudes of the real and imaginary 
parts are conveyed through intensity modulation (IM) of LEDs, the sign 
information is conveyed through spatial indexing of LEDs. Since four 
LEDs form one complex modulation unit, we term this as {\em quad-LED 
complex modulation (QCM)} \cite{qcm}. 
\item The second idea is to exploit the
representation of a complex symbol in polar coordinates. Instead of 
conveying the real and imaginary parts of a complex symbol 
and their sign information using four LEDs in QCM, we can convey only 
the magnitude and phase of a complex symbol. We need only two LEDs 
for this purpose and there is no sign information to convey in this 
representation. So we use only two LEDs to form a single modulation 
unit in this case. We term this scheme as {\em dual-LED complex 
modulation (DCM)} since two LEDs constitute one complex modulator. 
\item
The third proposed idea is to bring in the advantages of spatial 
modulation to the DCM scheme. Instead of using all the four LEDs to 
transmit one complex symbol (as in QCM), we choose two out of four LEDs 
to transmit the magnitude and phase of a complex symbol as in DCM scheme. 
Since we have to choose one pair of LEDs (one BLOCK) out of two and each 
pair will perform the same operation as in DCM scheme, we term this scheme 
as {\em spatial modulation-DCM (SM-DCM)} \cite{thesis}.
\end{itemize}

We investigate the bit error performance of the proposed QCM, DCM, and
SM-DCM schemes\footnote{We note that in all the three schemes, the 
number of LEDs that will be simultaneously ON in a channel use is two.} 
through analysis and simulations. We obtain upper bounds 
on the bit error rate (BER) of QCM, DCM, and SM-DCM. These analytical 
bounds are very tight at high signal-to-noise ratios (SNR). Therefore, 
these bounds enable us to easily compute and plot the achievable rate 
contours for a desired a target BER (e.g., $10^{-5}$ BER) in QCM, DCM, 
and SM-DCM. The analytical and simulation results show that the QCM, DCM, 
SM-DCM schemes achieve good BER performance. DCM has the advantage of 
fewer LEDs (2 LEDs) per complex modulator and better performance compared 
to QCM for small-sized modulation alphabets (e.g., 8-QAM). On the other 
hand, QCM has the advantage of additional degrees of freedom (4 LEDs) 
compared to DCM, because of which it achieves better performance compared 
to DCM for large alphabet sizes (e.g., 16-QAM, 32-QAM, 64-QAM). SM-DCM 
achieves better performance compared to DCM and QCM for small-sized 
modulation alphabets (e.g., 16-QAM) since it requires smaller modulation 
size. On the other hand, for large alphabet sizes, SM-DCM performs better 
compared to QCM at low $E_b/N_0$ values due to lower order modulation 
size, whereas at high $E_b/N_0$ values, SM-DCM performance degrades 
because of the reduced average relative distance between transmit vectors 
compared to QCM.

Since QCM and DCM can directly handle complex symbols
in VLC, techniques which are applied to complex modulation schemes 
to improve performance in RF wireless channels can be applied to VLC 
as well. For example, it is known that rotation of complex 
modulation symbols can improve BER performance in RF wireless 
communication \cite{tse}. Motivated by this observation, we explore 
the possibility of achieving performance improvement in VLC through 
phase rotation of complex modulation symbols prior to mapping the 
signals to the LEDs in QCM. We term this scheme as QCM with phase 
rotation (QCM-PR). Results show that phase rotation of modulation
symbols indeed can improve the BER performance of QCM in VLC. We 
also study the proposed QCM and DCM schemes when used along with 
OFDM; we refer to these schemes as QCM-OFDM and DCM-OFDM. We present 
zero-forcing and minimum distance detectors and their performance
for QCM-OFDM and DCM-OFDM. 

The rest of this paper is organized as follows. The indoor VLC system 
model is presented in Section \ref{sec2}. The proposed QCM, 
QCM-PR, and QCM-OFDM schemes and their performance are presented in 
Section \ref{sec3}. Section \ref{sec4} presents the proposed DCM and 
DCM-OFDM schemes and their performance. Section \ref{sec5} presents 
the proposed SM-DCM scheme and it performance. Section \ref{sec6} 
presents the spatial distribution of the received SNRs and the rate
contours achieved in QCM, DCM, and SM-DCM. Conclusions are presented in 
Section \ref{sec7}.

\section{Indoor VLC system model}
\label{sec2}
\begin{figure}
\centering
\includegraphics[height=1.275in]{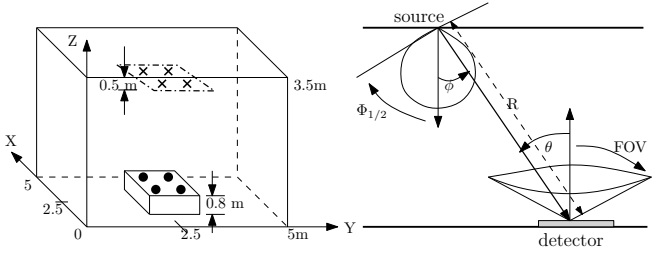}
\caption{Geometric set-up of the considered indoor VLC system.
A dot represents a photo detector and a cross represents an LED.}
\label{sys}
\vspace{-1mm}
\end{figure}

Consider an indoor VLC system with $N_t$ LEDs (transmitter) and $N_r$
photo detectors (receiver). Assume that the LEDs have a Lambertian
radiation pattern \cite{chann2},\cite{new1}. In a given channel use,
each LED is either OFF or emits light with some intensity which
is the magnitude of either the real part or imaginary part of a
complex modulation symbol. An LED which is OFF implies a light 
intensity of zero. Let $\vx = [x_1 \ x_2 \ \cdots x_{N_t}]^T$ denote 
the $N_t\times 1$ transmit signal vector, where $x_i$ is the light 
intensity emitted by the $i$th LED. Let $\mh$ denote the 
$N_r\times N_t$ MIMO VLC channel matrix:
\begin{eqnarray}
\mh=
\begin{bmatrix}
h_{11} & h_{12} & h_{13} & \cdots & h_{1N_t} \\
h_{21} & h_{22} & h_{23} & \cdots & h_{2N_t} \\
\vdots & \vdots & \ddots & \vdots & \vdots \\
h_{N_r1}& h_{N_r2} & h_{N_r3} & \cdots & h_{N_rN_t}
\end{bmatrix},
\end{eqnarray}
where $h_{ij}$ is the channel gain between $j$th LED and $i$th photo 
detector, $j=1,2,\cdots,N_t$ and $i=1,2,\cdots,N_r$. As in \cite{vlc5}, 
we consider only the line-of-sight (LOS) paths between the LEDs and the 
photo detectors. From \cite{chann2}, the LOS channel gain $h_{ij}$ is 
calculated as (see Fig. \ref{sys} for the definition of various angles 
in the model)
\begin{equation}
{h_{ij}} = \frac{n+1}{2\pi}\cos^{n}{\phi_{ij}}\,
\cos{\theta_{ij}}\frac{A}{R_{ij}^2}
\mbox{rect}\Big(\frac{\theta_{ij}}{FOV}\Big),
\label{channel}
\end{equation}
where $\phi_{ij}$ is the angle of emergence with respect to the $j$th
source (LED) and the normal at the source, $n$ is the mode number of 
the radiating lobe given by
$n=\frac{-\ln(2)}{\ln\cos{\Phi_{\frac{1}{2}}}},$
$\Phi_\frac{1}{2}$ is the half-power semiangle of the LED \cite{new1},
$\theta_{ij}$ is the angle of incidence at the $i$th photo detector,
$A$ is the area of the detector, $R_{ij}$ is the distance between the
$j$th source and the $i$th detector, FOV is the field of view of the
detector, and $\mbox{rect}(x)=1$, if $|x|\leq 1$, and  
$\mbox{rect}(x)=0$, if $|x|> 1$. 

The LEDs and the photo detectors are placed in a room of size
5m$\times$5m$\times$3.5m as shown in Fig. \ref{sys}. The LEDs are
placed at a height of 0.5m below the ceiling and the photo detectors
are placed on a table of height 0.8m. Let $d_{tx}$ denote the distance
between the LEDs and $d_{rx}$ denote the distance between the photo
detectors. 

Assuming perfect synchronization, the $N_r\times 1$ received signal vector
at the receiver is given by
\begin{eqnarray}
\vy = a\mh\vx+\vn,
\label{sysmodel}
\end{eqnarray}
where $a$ is the responsivity of the detector \cite{new2} and $\vn$ is
the noise vector of dimension $N_r\times 1$. Each element in the noise
vector $\vn$ is the sum of received thermal noise and ambient shot light
noise, which can be modeled as i.i.d. real AWGN with zero mean and variance
$\sigma^2$ \cite{chann1}. 
The average received signal-to-noise ratio (SNR) is given by
${\overline{\gamma}}=\frac{a^2P_r^2}{\sigma^2}$,
where $P_r^2 = \frac{1}{N_r}\sum\limits_{i=1}^{N_r}\se [{|\mh_i\vx|}^2]$,
and $\mh_i$ is the $i$th row of $\mh$.

\section{Proposed QCM scheme}
\label{sec3}

\subsection{QCM transmitter}
\label{sec3a}
\begin{figure}[t]
\centering
\includegraphics[width=3.4in, height=1.30in]{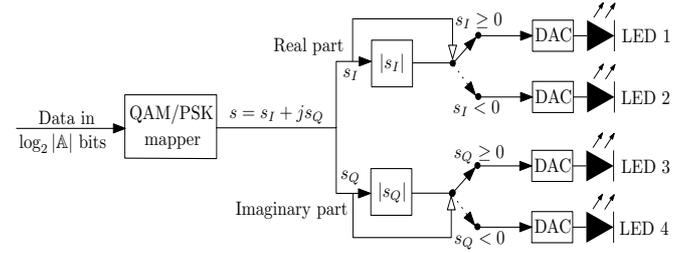}
\caption{QCM transmitter.}
\label{fig3}
\vspace{1mm}
\end{figure}

The proposed QCM scheme uses four LEDs at the transmitter. Figure
\ref{fig3} shows the block diagram of a QCM transmitter. Let 
$\mathbb A$ denote the complex modulation alphabet used (e.g., 
QAM). In each channel use, one complex symbol from $\mathbb A$ 
(chosen based on $\log_2|\mathbb A|$ information bits) is signaled 
by the four LEDs as described below. 

Each complex modulation symbol can have a positive or negative real part, 
and a positive or negative imaginary part. For example, the signal set 
for 16-QAM is $\{\pm 1 \pm \mbox{j} 1, \ \pm 1 \pm \mbox{j} 3, \ 
\pm 3 \pm \mbox{j} 1, \ \pm 3 \pm \mbox{j} 3\}$. Let $s \in \mathbb A$ 
be the complex symbol to be signaled in a given channel use. Let 
\[
s = s_I + \mbox{j}s_Q,
\] 
where $s_I$ and $s_Q$ are the real and imaginary parts of $s$, 
respectively. Two LEDs (say, LED1 and LED2) are used to convey the 
magnitude and sign of $s_I$ as follows. LED1 will emit with intensity 
$|s_I|$ if $s_I$ is positive $(\geq 0)$, whereas LED2 will emit with 
the same intensity $|s_I|$ if $s_I$ is negative $(< 0)$. Note that, 
since $s_I$ is either $\geq 0$ or $< 0$, only any one of LED1 and LED2 
will be ON in a given channel use and the other will be OFF. In a similar 
way, the remaining two LEDs (i.e., LED3 and LED4) will convey the magnitude 
and sign of $s_Q$ in such a way that LED3 will emit intensity $|s_Q|$ 
if $s_Q$ is $\geq 0$, whereas LED4 will emit with the same intensity 
$|s_Q|$ if $s_Q$ is $< 0$. Therefore, QCM sends one complex symbol in 
one channel use. The mapping of the magnitudes and signs of $s_I$ and 
$s_Q$ to the activity of LEDs in a given channel use is summarized in 
Table \ref{tab}. 

\begin{table}[t]
\centering
\begin{tabular}{|c||l||c|l|}
\hline
Real part & Status of LEDs & Imag. part & Status of LEDs \\
$s_I$ & & $s_Q$ & \\ \hline \hline
$\geq 0$ & LED1 emits $|s_I|$ & $\geq 0$ & LED3 emits $|s_Q|$  \\ 
& LED2 is OFF   &              & LED4 is OFF \\ \hline
$< 0$ & LED1 is OFF & $< 0$ & LED3 is OFF \\ 
& LED2 emits $|s_I|$ & & LED4 emits $|s_Q|$ \\ \hline
\end{tabular}
\vspace{2mm}
\caption{\label{tab} Mapping of complex symbol $s$ (with real part $s_I$ 
and imaginary part $s_Q$) to LEDs activity in QCM.}
\vspace{-8mm}
\end{table}

{\em Example 1:} If $s=-3+\mbox{j}1$, then the LEDs will be activated 
as follows: LED1: OFF; LED2: emits 3; LED3: emits 1; LED4: OFF. The 
$N_t\times 1$ (i.e., $4\times 1$) QCM transmit vector in this example 
is ${\bf x}=[0 \ 3 \ 1 \ 0]^T$. Likewise, if $s=1-\mbox{j}3$, then 
activation of LEDs will be as follows: LED1: emits 1; LED2: OFF; 
LED3: OFF; LED4: emits 3. The corresponding QCM transmit vector is 
${\bf x}=[1 \ 0 \ 0 \ 3]^T$.  

{\em Remark 1:} Because of the proposed mapping, in any given 
channel use, two LEDs (one among LED1 and LED2, and another one 
among LED3 and LED4) will be ON simultaneously and the remaining 
two LEDs will be OFF. 

{\em Remark 2:} The complex symbol conveyed in a channel use can be 
detected from the received QCM signal using the 
knowledge of the QCM map (Table \ref{tab}) at the receiver. 

\vspace{-2mm}
\subsection{QCM signal detection}
\label{sec3b}
\begin{figure}[t]
\centering
\includegraphics[width=2.75in, height=1.5in]{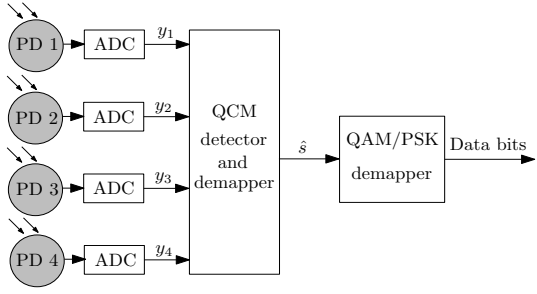}
\caption{QCM receiver.}
\label{fig4}
\vspace{-2mm}
\end{figure}

Figure \ref{fig4} shows the block diagram of a QCM receiver with $N_r=4$
PDs. Following the system model in Sec. \ref{sec2}, the $N_r\times 1$ 
received signal vector at the output of the PDs is given by (\ref{sysmodel}).
Assuming perfect channel knowledge at the receiver, the maximum likelihood
(ML) estimate of the transmit vector ${\bf x}$ is obtained as
\begin{equation}
\hat{\bf x}_{ML}=\argmin_{{\bf x}\in {\mathbb S}_Q} \|{\bf y}-a{\bf Hx}\|^2,
\end{equation}
where ${\mathbb S}_Q$ denotes the QCM signal set (consisting of all 
possible ${\bf x}$ vectors). The detected vector $\hat{\bf x}_{ML}$ is 
demapped to the corresponding complex symbol $\hat{s}_{ML}$, which is 
then demapped to get the corresponding information bits.

\begin{figure}[h]
\centering
\includegraphics[height=1.25in]{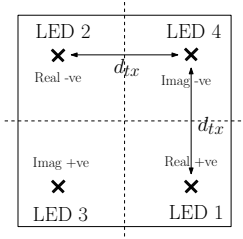}
\vspace{-1mm}
\caption{Placement of LEDs and signal mapping to LEDs.}
\label{fig5}
\vspace{-3mm}
\end{figure}

\vspace{-2mm}
\subsection{BER performance of QCM}
\label{sec3c}
In this subsection, we present the BER performance of QCM obtained through 
an analytical upper bound on the BER and simulations.

\subsubsection{Upper bound on BER}
\label{ub_qcm}
Consider the QCM system model in (\ref{sysmodel}). Normalizing the 
elements of the noise vector to unit variance, the received vector 
can be written as 
\begin{eqnarray}
\vy = \frac{a}{\sigma}\mh\vx +\vn.
\label{rec}
\end{eqnarray}
The ML detection rule for QCM can be rewritten as
\begin{equation}
\hat{\bf x}_{ML} = \argmin_{\bf x} \big(\frac{a}{\sigma}\|\mh\vx\|^2-2\vy ^T\mh\vx\big).
\label{xhat}
\end{equation}
Assuming that the channel matrix $\mh$ is known at the receiver, the
pairwise error probability (PEP) -- probability that the receiver decides
in favor of the signal vector $\vx _2$ when $\vx _1$ was transmitted --
can be written as
\begin{eqnarray}
\hspace{-9mm}
PEP({\vx _1} \rightarrow {\vx_2}|\mh) & = & 
\mbox{Pr}({\vx _1} \rightarrow {\vx_2}|\mh) \nonumber\\
&\hspace{-55mm}=&\hspace{-30mm}\mbox{Pr}\bigg (\vy ^T\mh({\vx _2}-{\vx _1})> \frac{a}{2\sigma}\big(\|\mh{\vx _2}\|^2-\|\mh{\vx _1}\|^2\big )\bigg) \nonumber \\
 &\hspace{-55mm}=&\hspace{-30mm}\mbox{Pr}\bigg (\frac{2\sigma}{a}\vn^T\mh(\vx_2 -\vx_1)>\|\mh(\vx_2 -\vx_1 )\|^2\bigg).
\label{PEP1}
\end{eqnarray}
Define $z\Define\frac{2\sigma}{a}\vn^T\mh(\vx_2 -\vx_1)$. We can see
that $z$ is a Gaussian random variable with mean $\se (z)=0$ and variance
$\mbox{Var}(z)=\frac{4\sigma^2}{a^2}\|\mh({\vx _2}-{\vx _1})\|^2$.
Therefore, (\ref{PEP1}) can be written as
\begin{equation}
PEP({\vx _1} \rightarrow {\vx_2}|\mh)
=Q\bigg (\frac{a}{2\sigma}\|\mh({\vx _2}-{\vx _1})\|\bigg).
\label{PEPQ}
\end{equation}
Let $\eta = \log_2{M}$ bpcu and $L\Define|{\mathbb S}_Q|$. 
An upper bound on the QCM BER for ML detection can be obtained using 
union bound as
\begin{eqnarray}
\hspace{-5mm}BER &\hspace{-2mm}\leq&\hspace{-2mm}
\frac{1}{L}\sum_{i=1}^{L}\sum_{j=1,i\neq j}^{L-1}PEP({\vx _i}\rightarrow {\vx _j}|\mh)\frac{d_{H}({\vx _i},{\vx _j})}{\eta} \nonumber \\
&\hspace{-2mm}=&\hspace{-2mm}\frac{1}{L}\sum_{i=1}^{L}\sum_{j=1,i\neq j}^{L-1}\hspace{-3mm}Q\bigg (\frac{a}{2\sigma}\|\mh({\vx_j}-{\vx_i})\|\bigg)\frac{d_{H}({\vx _i},{\vx _j})}{\eta},
\label{BER1}
\end{eqnarray}
where ${d_{H}({\vx _i},{\vx _j})}$ is the Hamming distance between the
bit mappings corresponding to the signal vectors $\vx _i$ and $\vx _j$.

\subsubsection{Results and discussions}
We evaluated the BER performance of QCM through the analytical upper 
bound and simulations. The various system parameters used in the 
performance evaluation are listed in Table \ref{tab1}. The placement 
of LEDs and the signal mapping to these LEDs used are shown 
Fig. \ref{fig5}. We evaluate the performance of QCM for various 
modulation alphabets including BPSK, 4-, 16-, and 64-QAM. 

\begin{table}[t]
\centering
\begin{tabular}{|l||l|l|}
\hline
        & Length $(X)$  & 5m \\ \cline{2-3}
Room    & Width ($Y$)   & 5m \\ \cline{2-3}
        & Height ($Z$)  & 3.5m  \\ \hline \hline
        & No. of LEDs ($N_t$) & 4 \\ \cline{2-3}
        & Height from the floor & 3m \\ \cline{2-3}
        & Elevation     & $-90\degree$ \\ \cline{2-3}
Transmitter & Azimuth   & $0\degree$ \\ \cline{2-3}
        & $\Phi_{1/2}$  & $60\degree$ \\ \cline{2-3}
        & Mode number, $n$ & 1 \\ \cline{2-3}
        & $d_{tx}$      & 0.2m to 4.8m \\ \hline \hline
        & No. of PDs ($N_r$) & 4 \\ \cline{2-3}
        & Height from the floor & 0.8m\\ \cline{2-3}
        & Elevation     & $90\degree$ \\ \cline{2-3}
Receiver & Azimuth      & $0\degree$ \\ \cline{2-3}
        & Responsivity, $a$ & 1 Ampere/Watt  \\ \cline{2-3}
        & FOV           & $85\degree$ \\ \cline{2-3}
        & $d_{rx}$      & 0.1m \\ \hline
\end{tabular}
\vspace{2mm}
\caption{\label{tab1} System parameters in the considered indoor VLC system.}
\vspace{-6mm}
\end{table}

\begin{figure}[t]
\centering
\includegraphics[width=3.5in, height=2.50in]{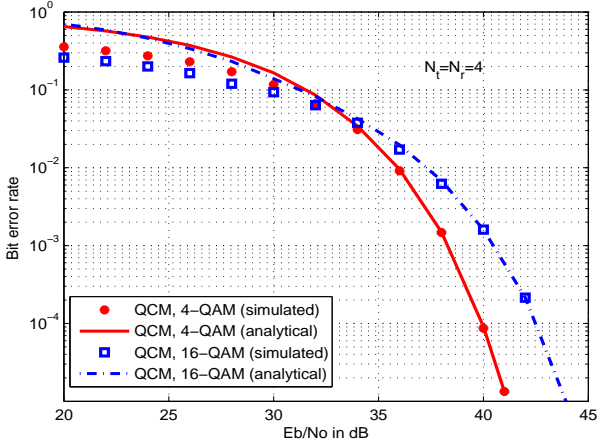}
\vspace{-7mm}
\caption{Comparison of analytical upper bound and simulated BER
for QCM with $d_{tx}=1$m, 4-QAM, and 16-QAM.} 
\label{fig17}
\vspace{-5mm}
\end{figure}

\begin{figure}
\centering
\includegraphics[width=3.5in, height=2.5in]{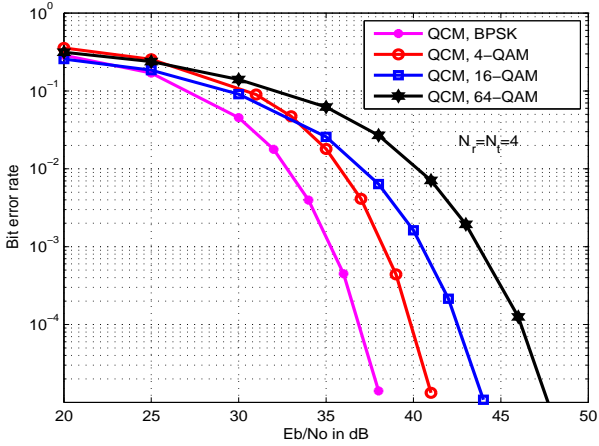}
\vspace{-7mm}
\caption{BER performance of QCM with BPSK, 4-QAM, 16-QAM, and 64-QAM
at $d_{tx}=1$m.} 
\label{fig6}
\vspace{-4mm}
\end{figure}

In Fig. \ref{fig17}, we plot the analytical upper bound and the
simulated BER for QCM with $d_{tx}=1$m, 4-QAM, 16-QAM, $N_r=4$, and 
ML detection. It can be seen that the analytical upper bound on the 
BER of QCM is very tight at moderate to high SNRs. In Fig. \ref{fig6}, 
we plot the simulated BER of QCM with $d_{tx}=1$m and ML detection for 
BPSK (1 bpcu), 4-QAM (2 bpcu), 16-QAM (4 bpcu), and 64-QAM (6 bpcu). 
From Fig. \ref{fig6}, we observe that QCM achieves $10^{-4}$ BER at 
an $E_b/N_0$ of about 37 dB for BPSK, 40 dB for 4-QAM, 42.5 dB for 
16-QAM, and 46.5 dB for 64-QAM. This observed increase in the required 
$E_b/N_0$ for increased QAM size is because of the reduced minimum 
distance for larger QAM size, and it is in line with what happens in 
conventional RF modulation. In addition, we observe crossovers which 
show better performance for larger-sized QAM at low SNRs (e.g., crossover 
between the performance of 4-QAM and 16-QAM at around $4\times 10^{-2}$  
BER). This crossover occurs due to the degrading effect of an equal-power 
interferer\footnote{Signals from two active LEDs interfere with each other 
at the receiver.} on the one hand, and the benefit of a strong interferer 
in multiuser detection\footnote{A strong interferer can be effectively 
canceled in a multiuser detector \cite{verdu}.} on the other hand. This 
can be further explained with the following example. The signal received at 
the $i$th PD is $y_i=h_l|s_I|+h_k|s_Q|+n_i$, where $h_l$ and $h_k$ are 
the channel gains corresponding to the LEDs chosen to transmit $|s_I|$ 
and $|s_Q|$, respectively. For 4-QAM, the transmit signals from both 
the active LEDs will be 1 (i.e., both $|s_I|$ and $|s_Q|$ will be 1). 
Whereas for 16-QAM, the transmit signal from each active LED can be 
1 or 3 (i.e., $|s_I|$ can be 1 or 3, and so is $|s_Q|$). Therefore, 
the received signal for 4-QAM is $y_i=h_l+h_k+n_i$. Also, $h_l$ and 
$h_k$ can be nearly equal because of high channel correlation, making 
4-QAM detection unreliable at low SNRs. Whereas, since 
$|s_I|,|s_Q|\in\{1,3\}$ in 16-QAM, the effect of channel correlation 
between $h_l$ and $h_k$ in 16-QAM detection is reduced. That is, 
$\se\big(\big|h_l|s_I|-h_k|s_Q|\big|\big)$ is larger for 16-QAM 
compared to that for 4-QAM.

\begin{figure}
\centering
\includegraphics[width=3.5in, height=2.50in]{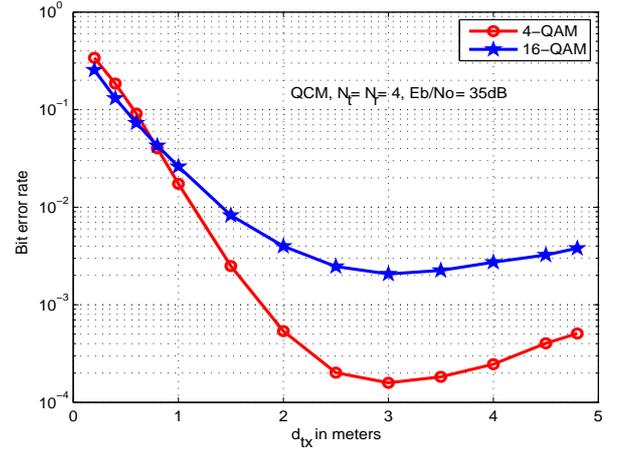}
\vspace{-6mm}
\caption{BER performance of QCM as a function of $d_{tx}$ for 4-QAM
and 16-QAM at $E_b/N_0=35$ dB.} 
\label{fig7}
\vspace{-4mm}
\end{figure}

{\em Effect of varying $d_{tx}$:}
Figure \ref{fig7} shows the effect of varying the spacing between the
LEDs ($d_{tx}$ varied in the range 0.2m to 4.8m) on the BER performance 
of QCM with 4-QAM and 16-QAM at $E_b/N_0=35$ dB. From Fig. \ref{fig7},
we see that there is an optimum $d_{tx}$ (around 3m) which gives the 
best BER performance. If $d_{tx}$ is increased above and decreased 
below this optimum spacing, the BER worsens. The reason for this 
optimum can be explained as follows. On the one hand, the channel 
gains get weaker as $d_{tx}$ is increased. This reduces the received 
signal level, which is a source of BER degradation. On the other hand, 
the channel correlation also gets weaker as $d_{tx}$ is increased. 
This reduced channel correlation is a source of BER improvement. 
These opposing effects of weak channel gains and weak channel 
correlations for increasing $d_{tx}$ results in an optimum spacing.
Also, as observed and explained in Fig. \ref{fig6}, in Fig. \ref{fig7} 
also we see that QCM with 16-QAM can perform a little better than QCM 
with 4-QAM when $d_{tx}$ is small and channel correlation is high.

\subsection{QCM with phase rotation}
\label{sec3d}
As we mentioned earlier, rotation of complex modulation symbols is known 
to improve BER performance in RF communications \cite{tse}. Motivated by 
this and the fact that QCM allows the use of complex modulation alphabets 
in VLC, in this subsection we study the performance QCM scheme with 
rotation of the complex modulation symbols. 

\subsubsection{QCM-PR transmitter}
\label{sec3d1}
In QCM with phase rotation, a complex symbol from a modulation alphabet
$\mathbb A$ is rotated by a phase angle of $\theta$ before being 
transmitted by the quad-LED setup. Let $s\in{\mathbb A}$ be the 
complex symbol chosen based on the input information bits. Instead
of sending the symbol $s$ as such in QCM, the QCM-PR transmitter 
sends the rotated complex symbol $s'$ given by
\begin{equation}
s'=e^{j\theta}s
\end{equation} 
through the quad-LED setup as described before. Therefore, 
\begin{eqnarray}
s'_I&=&s_I\cos\theta-s_Q\sin\theta, \nonumber \\
s'_Q&=&s_I\sin\theta+s_Q\cos\theta.
\end{eqnarray}
Let $\vx'$ be the QCM transmit vector constructed using $s'_I$ and $s'_Q$.
Now, $\vx'$ is the transmitted vector corresponding to the 
complex signal $s$ rotated by a phase angle $\theta$.

\subsubsection{QCM-PR signal detection}
\label{sec3d2}
We assume that the angle of rotation $\theta$ is known both at the transmitter
and receiver. The ML estimate of the transmitted symbol $s$ is then given by
\begin{equation}
\hat{\bf x}'_{ML}=\argmin_{{\bf x}'\in {\mathbb S}_{QP}} \|{\bf y}-a{\bf Hx'}\|^2.
\end{equation}
where ${\mathbb S}_{QP}$ denotes the QCM-PR signal set. The detected 
vector $\hat{\bf x}'_{ML}$ is demapped to the corresponding complex 
symbol $\hat{s}'_{ML}$, which is then demapped to get the corresponding 
information bits.

\subsubsection{BER performance of QCM-PR}
\label{sec3d3}
We evaluated the BER performance of QCM-PR scheme. The simulation 
parameter settings, LEDs placement, and signal mapping to LEDs are 
same as those used in Sec. \ref{sec3c}. 

\begin{figure}
\centering
\includegraphics[width=3.5in, height=2.5in]{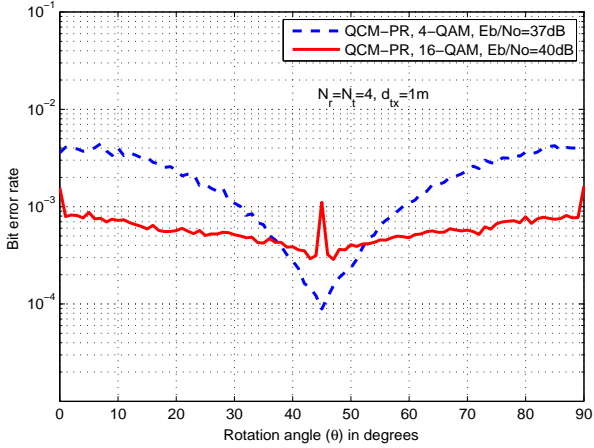}
\vspace{-6mm}
\caption{BER performance of QCM-PR as a function of rotation angle
$\theta$ for 4-QAM, $E_b/N_0=37$ dB and 16-QAM, $E_b/N_0=40$ dB
at $d_{tx}=1$m.}
\label{fig8}
\vspace{-4mm}
\end{figure}

\begin{figure}
\centering
\includegraphics[width=3.5in, height=2.5in]{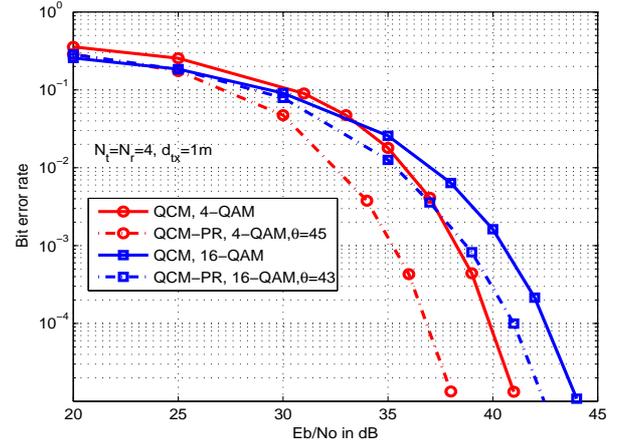}
\vspace{-6mm}
\caption{BER versus $E_b/N_0$ characteristics of QCM and QCM-PR 
for 4-QAM and 16-QAM at $d_{tx}=1$m.}
\label{fig9}
\vspace{-4mm}
\end{figure}

{\em BER as a function of rotation angle $\theta$:}
In Fig. \ref{fig8}, we plot the BER of QCM-PR scheme as a function of 
the rotation angle $\theta$ (in degrees) at $d_{tx}=1$m. BER plots for 
4-QAM with $E_b/N_0=37$ dB and 16-QAM with $E_b/N_0=40$ dB are shown. 
We limit the range of $\theta$ value in the x-axis from $0^\circ$ to 
$90^\circ$ as the pattern of the plots repeat after $90^\circ$ due to 
symmetry. Note that $\theta=0^\circ$ corresponds to the basic QCM without 
rotation. The following interesting observations can be made from
Fig. \ref{fig8}. First, for both 4-QAM and 16-QAM, the BER plots are 
symmetrical with respect to $45^\circ$, which can be expected. 
Second, for 4-QAM, $\theta=45^\circ$ happens to be the optimum 
rotation which gives the best BER\footnote{It is interesting to note 
that QCM-PR with 4-QAM and $\theta=45^\circ$ rotation specializes to 
SSK with $N_t=4$. That is, the 4-QAM signal set when rotated
by $45^\circ$ becomes $\{1+\mbox{j}0,0+\mbox{j}1,-1+\mbox{j}0,
0-\mbox{j}1\}$ When mapped to the LEDs as per QCM, the resulting
QCM signal set becomes $\{[1 0 0 0]^T, [0 0 1 0]^T, [0 1 0 0]^T,
[0 0 0 1]^T\}$, which is the same as the SSK signal set with
$N_t=4$. Because of this, only one LED will be ON at a time in 
QCM-PR with $\theta=45^\circ$ and therefore there will be no 
interference.}. Note that there is more than an order improvement in 
BER at this optimum rotation compared to basic QCM without rotation 
(see BERs at $\theta=0^\circ$ and $\theta=45^\circ$). Third, for 
16-QAM, there are two optimum angles around $45^\circ$ because of 
symmetry; $\theta=43^\circ$ is one of them. 
Figure \ref{fig9} shows a comparison between the BER performance
of QCM-PR (with optimum rotation angles) and QCM (no rotation) at 
$d_{tx}=1$m.  BER plots for 4-QAM and 16-QAM are shown. It can be 
seen that optimum phase rotation improves the BER performance by 
about 2 to 3 dB.

\begin{figure}[t]
\centering
\includegraphics[width=3.5in, height=2.5in]{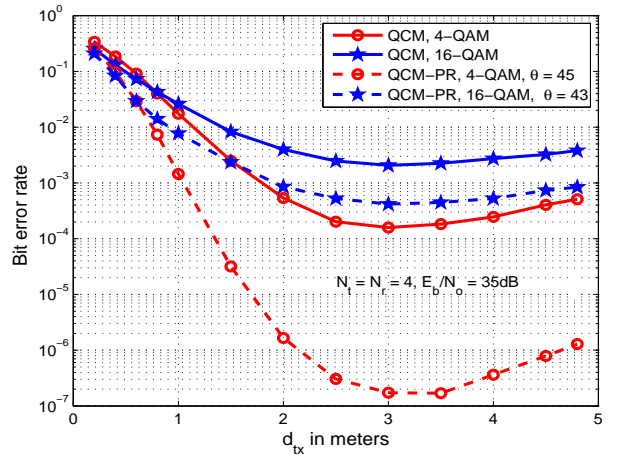}
\vspace{-6mm}
\caption{BER versus LED spacing ($d_{tx}$) characteristics of QCM and 
QCM-PR for 4-QAM and 16-QAM at $E_b/N_0=35$ dB.}
\label{fig10}
\vspace{-4mm}
\end{figure}

{\em QCM-PR vs QCM for different $d_{tx}$:}
Figure \ref{fig10} shows how varying $d_{tx}$ affects the BER performance 
of QCM-PR and QCM at $E_b/N_0=35$ dB. As observed for QCM in Fig. \ref{fig7},  
we see that there is an optimum spacing in QCM-PR as well, which is due 
to the opposing effects of weak channel gains and weak channel correlation
for increasing $d_{tx}$ values. QCM-PR achieves better performance compared 
to QCM. For example, at $d_{tx}=3$m, there is about 3 orders of BER 
improvement for 4-QAM. This reinforces the benefit of phase rotation.

\subsection{QCM-OFDM}
\label{sec3e}
Since QCM allows the transmission of complex symbols using the quad-LED
setup, OFDM signaling can be carried out using QCM. In this subsection, 
we present the QCM-OFDM scheme, its detection, and performance.

\subsubsection{QCM-OFDM transmitter}
\label{sec3e1}
In the QCM-OFDM transmitter, $N$ complex symbols from $\mathbb A$ 
(chosen based on $N\log_2|\mathbb A|$ information bits) will be 
transmitted by the four LEDs in $N$ channel uses, where $N$ is 
the number of subcarriers. The $N$ complex symbols 
$\vv=[v_1, v_2, \cdots, v_N]^T\in{\mathbb A}^N$ are transformed 
using inverse Fourier transform (IFFT) to obtain the complex 
transmit symbols $\vs=[s_1, s_2, \cdots, s_N]^T=\mf^H\vv$, where
$\mf$ is the Fourier transform matrix. The $N$ output symbols from
the IFFT block are then transmitted one by one in $N$ channel uses 
by the quad-LED setup in the QCM transmitter. Thus, effectively $N$ 
complex modulation symbols are sent in $N$ channel uses. Let 
${\bf x}_n$ denote the $N_t\times 1$ (i.e., $4\times 1$) transmit 
vector corresponding to $s_n$, $n=1,2,\cdots,N$. 

\subsubsection{QCM-OFDM signal detection}
\label{sec3e2}
Let $\my=[\vy_1, \vy_2, \cdots, \vy_N]$ be the matrix of received 
vectors corresponding to the matrix of transmit vectors 
${\bf X}=[{\bf x}_1, {\bf x}_2, \cdots, {\bf x}_N]$, i.e., 
corresponding to the signal
vector $\vs=[s_1, s_2, \cdots, s_N]^T$. Before performing Fourier 
transform (FFT) operation, we need to detect the transmitted symbols
$s_i\in[0,\infty)$. This detection involves two stages, namely, ($i$) 
active LEDs identification, and ($ii$) complex symbol reconstruction.

{\em Active LEDs identification:}
To discern the two active LEDs in the quad-LED setup, we compute 
\begin{equation}
z_{i,j}=(\vh_j^T\vh_j)^{-1}\vh_j^T\vy_i, \ \ j=1,2,3,4, \ \ i=1,\cdots,N,
\end{equation}
where $\vh_j$ is the $j$th column of the channel matrix ${\bf H}$. For 
the $i$th channel use, the LEDs corresponding to the two largest values 
of $|z_{i,j}|$ are detected to be active. That is, if $i_1$ and $i_2$ 
are the indices of the active LEDs in the $i$th channel use, then
\begin{eqnarray*}
\hat i_1 &=& \argmax_{j\in\{1,2,3,4\}} |z_{i,j}| \quad \quad i_1\in\{1,2,3,4\}\\
\hat i_2 &=& \argmax_{j\in\{1,2,3,4\}\setminus i_1} |z_{i,j}| \quad \quad i_2\in\{1,2,3,4\}\setminus i_1.
\end{eqnarray*}

{\em Complex symbol reconstruction:}
After identifying the active LEDs, we need to detect $s_I$ and $s_Q$.
This can be achieved through a zero-forcing (ZF) type detector. Let
$\vs_i=[s_I^i, s_Q^i]^T$ be the transmitted signal values corresponding 
to the complex signal $s_i$. Form $\mh_{ZF}$ matrix using the 
$i_1$th and $i_2$th columns of ${\bf H}$ as 
$\mh_{ZF}=[\vh_{i_1} \, \vh_{i_2}]$. Now, the ZF detector output is 
given by 
\begin{equation}
{\hat \vs_i}=\frac{1}{r}(\mh_{ZF}^T\mh_{ZF})^{-1}\mh_{ZF}^T\vy_i.
\end{equation}
Finally, an estimate of the transmitted complex symbol is obtained as
${\hat s_i}={\hat s_I^i}+\mbox{j}{\hat s_Q^i}$. Now, 
${\hat \vv}=\mf{\hat \vs}$. The $N\log_2|\mathbb A|$ information bits 
are demapped from ${\hat \vv}$.

\subsubsection{Minimum distance detector}
\label{sec3e3}
The above zero forcing detector is a sub-optimal detector. Therefore, to 
further improve the performance of QCM-OFDM, we use a minimum distance 
(MD) detector. This detector is described as follows. Let ${\mathbb S}_F$ 
be the set of all possible values the vector $\vs$ can take, i.e., 
$\vs\in{\mathbb S}_F$ and $|{\mathbb S}_F|=|{\mathbb A}^N|$.
$\vx_n$ is the QCM transmit vector in the $n$th channel use, 
$n=1,2,\cdots,N$, and $\mx=[\vx_1, \vx_2, \cdots, \vx_N]$ is the matrix 
of QCM transmit vectors for one QCM-OFDM symbol. Let ${\mathbb S}_{QO}$ 
be the set of all possible values of the matrix $\mx$, i.e., 
$\mx\in{\mathbb S}_{QO}$ and $|{\mathbb S}_{QO}|=|{\mathbb A}^N|$. 
Therefore, for every $\vv\in{\mathbb A}$ there exists a corresponding 
matrix $\mx\in{\mathbb S}_{QO}$ and vice versa. The estimate of $\vv$ 
in the MD detector is obtained as
\begin{equation}
{\hat \vv}=\argmin_{\mx\in{\mathbb S}_{QO}} \|\my-a\mh\mx\|.
\end{equation}
The $N\log_2|\mathbb A|$ information bits are demapped from ${\hat \vv}$.

\subsubsection{BER performance of QCM-OFDM}
\label{sec3e4}
\begin{figure}
\centering
\includegraphics[width=3.5in, height=2.50in]{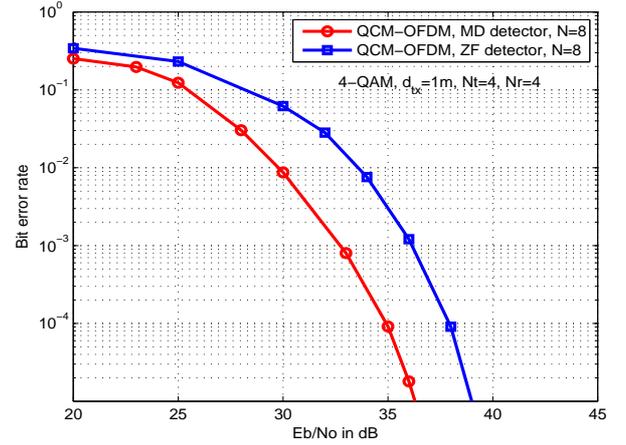}
\vspace{-6mm}
\caption{BER performance of QCM-OFDM with ZF detection and MD detection
for $N=8$, 4-QAM, $d_{tx}=1$m.}
\label{fig11}
\vspace{-4mm}
\end{figure}

\begin{figure}
\centering
\includegraphics[width=3.5in, height=2.50in]{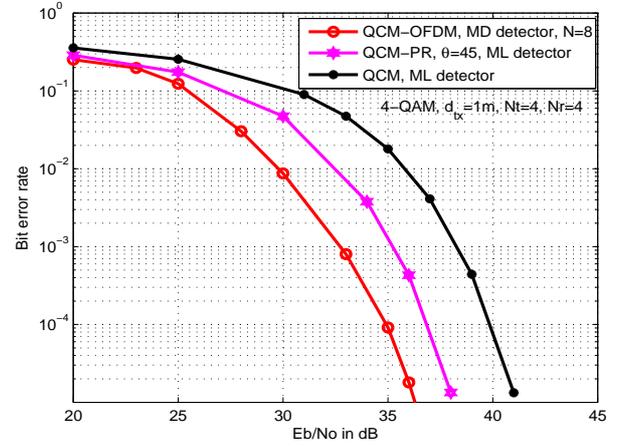}
\vspace{-6mm}
\caption{BER performance comparison between QCM, QCM-PR, and QCM-OFDM
for 4-QAM at $d_{tx}=1m$.}
\label{fig12}
\vspace{-4mm}
\end{figure}

We evaluated the BER performance of QCM-OFDM scheme. 
The simulation parameter settings, LEDs placement, and signal mapping
to LEDs are same as those used in Sec. \ref{sec3c}. Figure \ref{fig11} 
shows the BER performance of QCM-OFDM with $N=8$ and 4-QAM at $d_{tx}=1$m.
The performance achieved by ZF detection and MD detection (presented in
the previous subsection) are plotted. It can be seen that the MD detector
achieves better performance by 2.5 dB to 3.5 dB compared to the ZF detector.
In Fig. \ref{fig12}, we compare the performance of QCM, QCM-PR with optimum
rotation $\theta=45^\circ$, and QCM-OFDM with 4-QAM and $d_{tx}=1$m.  
It can be seen that QCM-OFDM with MD detection achieves better performance
compared to both QAM and QCM-PR. For example, at a BER of $10^{-4}$,
QCM-OFDM performs better than QCM-PR and QCM by about 2 dB and 5 dB,
respectively. 

\section{Proposed DCM scheme}
\label{sec4}
\begin{figure}[t]
\centering
\includegraphics[width=3.25in, height=1.00in]{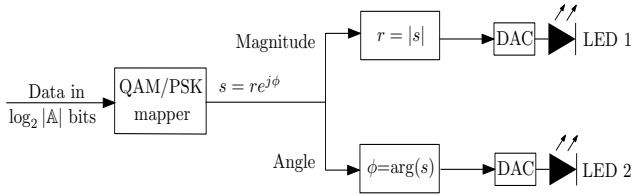}
\caption{DCM transmitter.}
\label{dcmtx}
\vspace{-4mm}
\end{figure}

The QCM scheme proposed in the previous section conveys the real and 
imaginary parts of a complex symbol and their sign information using
four LEDs. Representation of complex symbols in polar coordinates can be 
exploited instead. That is, it is adequate to convey only the magnitude 
and phase ($r,\phi$) of a complex symbol, for which only two LEDs
suffice and there is no sign information to convey. The DCM scheme 
proposed in this section is based on this idea. The proposed DCM scheme 
uses two LEDs to convey the magnitude and phase of a complex symbol. 
The block diagram of DCM transmitter is given in Fig. \ref{dcmtx}. The 
complex modulation symbol $s$ that is to be transmitted in a given channel 
use is split in to two real and non-negative parts, namely, the magnitude
of the complex symbol $r$ and the phase of the complex symbol $\phi$ such 
that
\begin{eqnarray}
s&=&re^{j\phi},\nonumber\\
r&=&|s|, \quad r\in {\mathbb R}^+, \nonumber \\  
\phi&=&\mbox{arg}(s), \quad \phi\in[0,2\pi).
\end{eqnarray}
Now, we use LED1 to transmit $r$ and LED2 to transmit $\phi$ through
intensity modulation. The $2\times 1$ DCM transmit vector $\vx$ is 
then given by $\vx=[r\quad \phi]^T$. 

{\em Example 2}: If the complex modulation symbol to be transmitted is 
$s=3+\mbox{j}3$, then $r=3\sqrt{2}$ and $\phi=\frac{\pi}{4}$. 
LED1 emits light of intensity $3\sqrt{2}$ and LED2 emits light of 
intensity $\frac{\pi}{4}$.

{\em Remark 3:} Note that in case of $M$-PSK modulation only the phases 
of the symbols convey information (because of constant magnitude). So 
the `magnitude-LED' (LED1) in DCM essentially becomes redundant, and 
only the angle value sent on the `angle-LED' (LED2) matters. Therefore, 
in this case, the DCM scheme can be viewed as equivalent to a single-LED 
scheme with $M$-PAM. However, in cases where $M$-PSK symbols undergo 
some pre-processing operation before transmission (e.g., IFFT operation 
on $M$-PSK symbols in OFDM), both the magnitude and angle of the resulting 
complex variables sent by LEDs 1 and 2 matter.

Now, the $N_r\times 1$ received signal vector at the receiver is given by
\begin{equation}
\vy=a\mh\vx+\vn,
\label{dcmeqn}
\end{equation}
and the ML estimate of the transmit vector ${\bf x}$ is given by 
\begin{equation}
\hat{\bf x}_{ML}=\argmin_{{\bf x}\in {\mathbb S}_D} \|{\bf y}-a{\bf Hx}\|^2,
\end{equation}
where ${\mathbb S}_D$ denotes the DCM signal set (consisting of all 
possible transmit vectors ${\bf x}$). The detected vector 
$\hat{\bf x}_{ML}$ is demapped to the corresponding complex symbol 
$\hat{s}_{ML}$, which is then demapped to get the corresponding 
information bits.

\subsection{BER performance of DCM}
\label{sec4a}
\begin{figure}
\centering
\includegraphics[width=3.5in, height=2.50in]{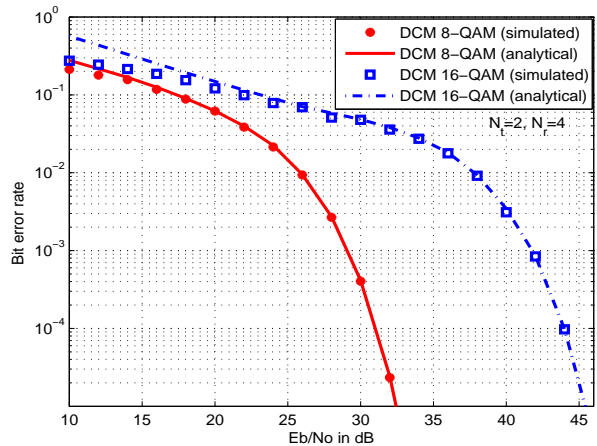}
\vspace{-6mm}
\caption{Comparison of analytical upper bound and simulated BER
for DCM with 8-QAM and 16-QAM.}
\label{fig18}
\vspace{-4mm}
\end{figure}

An upper bound on the BER of QCM for ML detection was obtained in Sec. 
\ref{ub_qcm}. In a similar way, an upper bound on the BER of DCM can 
be obtained. Figure \ref{fig18} shows the upper bound and simulated  
BER plots for DCM with 8-QAM and 16-QAM. The system parameters used 
are the same as in Table \ref{tab1}. The two LEDs are placed at the
locations of LED1 and LED2 specified in Fig. \ref{fig5}, and $d_{tx}=1$m. 
It can be seen that the upper bound is tight at moderate to high SNRs.

\begin{figure}
\centering
\includegraphics[width=3.5in, height=2.50in]{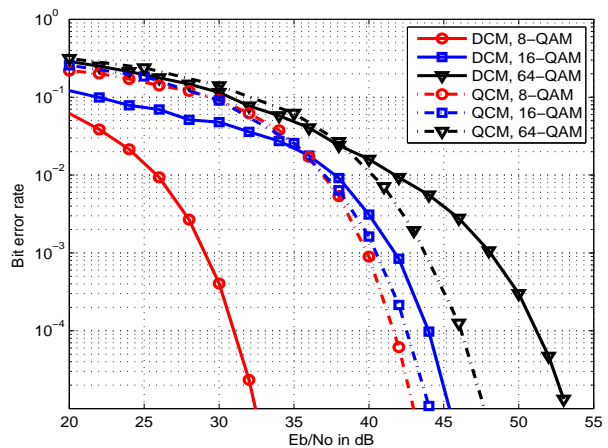}
\vspace{-6mm}
\caption{Comparison of the BER performance of DCM and QCM 
for 8-QAM, 16-QAM, and 64-QAM.}
\label{dcmber1}
\vspace{-4mm}
\end{figure}

Figure \ref{dcmber1} presents a BER vs $E_b/N_0$ performance comparison
between DCM and QCM for 8-QAM, 16-QAM, and 64-QAM using ML detection. 
Table \ref{dcmqcm} also presents a similar performance comparison 
between DCM, QCM, and QCM-PR. In this Table, we present the $E_b/N_0$
(in dB) required to achieve a BER of $10^{-3}$ in DCM, QCM, and 
QCM-PR\footnote{Note that, unlike in QCM, phase rotation
in DCM essentially gives the same performance as DCM without phase 
rotation. This is because $i$) the magnitude of the complex number 
does not change on rotation, and hence the values of $r$ transmitted 
by LED1 in DCM remain unaffected by rotation, and $ii)$ the relative 
distance between the intensity levels transmitted for the phase 
information $\phi$ by LED2 in DCM does not change on rotation.}
for 8-, 16-, 32-, 64-QAM. From Fig. \ref{dcmber1}, we observe that 
DCM achieves better performance compared to QCM for a
small-sized alphabet like 8-QAM; e.g., at a BER of $10^{-4}$, DCM 
performs better than QCM by about 10.4 dB for 8-QAM. On the other hand, 
for larger-sized alphabets like 16-QAM and 64-QAM, QCM outperforms DCM 
by about 1 dB and 5 dB, respectively, at $10^{-4}$ BER. Similar 
observations can be made in Table \ref{dcmqcm}; DCM requires a lesser 
$E_b/N_0$ to achieve $10^{-3}$ BER compared to QCM for 8-QAM, whereas
QCM requires less $E_b/N_0$ compared to DCM for 16-QAM, 32-QAM, and 
64-QAM. This is because, for large QAM sizes, the average relative 
distance between the transmit vectors in QCM is more compared to that 
in DCM, i.e., if ($\vx_1, \vx_2$) is a pair of vectors from the transmit
signal set, then $\mathbb{E}[\|\vx_1-\vx_2\|]$ is larger for the QCM 
signal set than for the DCM signal set. For example, the intensities 
emitted in QCM for 16-QAM could be one of $\{0, 1, 3\}$. Whereas, the 
intensities emitted in DCM for 16-QAM could be one among the 3 different 
possible values for $r$ and one among the 12 different possible values 
for $\phi$. As the size of the modulation alphabet ${\mathbb A}$ 
increases, the set of possible values of $r$ and $\phi$ increases as 
$O(|{\mathbb A}|)$ in DCM. Whereas, in QCM, as the size of the modulation 
alphabet increases, the cardinality of the set of possible transmit 
intensity levels is only $\sqrt{|{\mathbb A}|}/2+1$ if ${\mathbb A}$ 
is square QAM, and $\sqrt{|{\mathbb A}|/2}+1$ if ${\mathbb A}$ is 
non-square.

\begin{table}
\centering
\begin{tabular}{|c||c|c|c|}
\hline
Spectral efficiency & DCM & QCM & QCM-PR \\
in bpcu & & & \\
\hline\hline
3 (8-QAM)  & 29.2 dB & 39.8 dB & 39.2 dB \\ \hline
4 (16-QAM) & 41.8 dB & 40.6 dB & 38.6 dB \\ \hline
5 (32-QAM) & 45.5 dB & 41.8 dB & 40 dB \\ \hline
6 (64-QAM) & 48.2 dB & 43.7 dB & 40.2 dB \\ \hline
\end{tabular}
\vspace{3mm}
\caption{Comparison of $E_b/N_0$ required by DCM, QCM, and 
QCM-PR to achieve a BER of $10^{-3}$ with $M$-QAM alphabets.}
\label{dcmqcm}
\vspace{-6mm}
\end{table}

\subsection{DCM-OFDM}
\label{sec4b}
In this subsection, we present DCM-OFDM, its detection, and performance.  
In DCM-OFDM, complex OFDM symbols are transmitted through a dual-LED 
setup using intensity modulation. In DCM-OFDM transmitter, $N$ complex 
symbols from $\mathbb A$ carrying $N\log_2|\mathbb A|$ information 
bits are transformed by the IFFT matrix to get a $N\times 1$ vector of 
complex transmit symbols denoted as $\vs=[s_1, s_2, \cdots, s_N]^T=\mf^H\vv$, 
where $\vv=[v_1, v_2, \cdots, v_N]^T\in{\mathbb A}^N$. The elements of
$\vs$ are transmitted in $N$ channel uses through the DCM transmitter. 

\subsubsection{Zero-forcing based DCM-OFDM signal detection}
\label{sec4b1}
Let $\my=[\vy_1, \vy_2, \cdots, \vy_N]$ be the $N_r\times N$ matrix of 
received vectors corresponding to the $2\times N$ matrix of transmit 
vectors ${\bf X}=[{\bf x}_1, {\bf x}_2, \cdots, {\bf x}_N]$. We can 
detect the transmitted complex values at the receiver by performing a 
zero-forcing (ZF) equalization for the channel matrix $\mh$. That is, 
the transmitted vectors can be estimated as 
\begin{equation}
\widehat{\bf X}=\frac{1}{a}(\mh^T\mh)^{-1}\mh^T\my.
\label{dcmzf}
\end{equation}
The $N$ complex OFDM symbols can be reconstructed as
\begin{equation}
{\hat s_i}=\hat{x}_i(1)e^{\textrm{j}\hat{x}_i(2)},
\end{equation}
where $\hat{x}_i(k)$ is the $k$th element of the $i$th column vector in
the matrix $\widehat{\bf X}$. Now, OFDM demodulation is preformed as
${\hat \vv}=\mf{\hat \vs}$ and the $N\log_2|\mathbb A|$ information bits 
are demapped from it. The complexity of ZF detector is $O(2N_rN)$.

\subsubsection{Minimum distance detector}
\label{sec4b2}
We saw in Sec. \ref{sec3e3} that the minimum distance detector offers 
performance improvement over ZF detector for QCM-OFDM. Similarly, the 
performance of DCM-OFDM can be improved by using the minimum distance 
detector. Let ${\mathbb S}_{DO}$ be the set of all possible values of 
the DCM-OFDM transmit matrix $\mx$, i.e., $\mx\in{\mathbb S}_{DO}$ and 
$|{\mathbb S}_{DO}|=|{\mathbb A}^N|$. Thus, every matrix in 
$\mx\in{\mathbb S}_{DO}$ corresponds to an $N\times 1$ complex vector 
$\vv\in{\mathbb A}$. In the MD detector, $\vv$ can be estimated as
\begin{equation}
{\hat \vv}=\argmin_{\mx\in{\mathbb S}_{DO}} \|\my-a\mh\mx\|.
\end{equation}
The $N\log_2|\mathbb A|$ information bits are demapped from ${\hat \vv}$.
The complexity of the MD detector is $O(|{\mathbb A}|^N)$. 

\subsubsection{BER performance of DCM-OFDM}
\label{sec4b3}
\begin{figure}
\centering
\includegraphics[width=3.5in, height=2.50in]{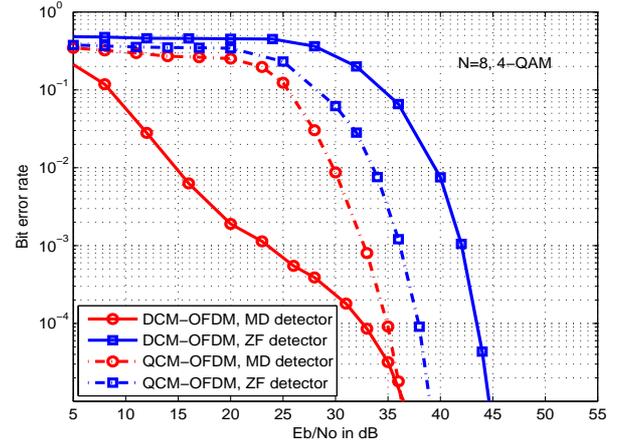}
\vspace{-6mm}
\caption{BER performance of DCM-OFDM with ZF and MD detectors for $N=8$
and 4-QAM.}
\label{dcmofdm}
\vspace{-4mm}
\end{figure}

The BER performance of DCM-OFDM scheme with $N=8$ and 4-QAM using ZF and 
MD detectors is presented in Fig. \ref{dcmofdm}. For comparison purpose, 
we have also plotted the performance of QCM-OFDM in Fig. \ref{dcmofdm}. 
The simulation parameter settings are same as those listed in Table 
\ref{tab1}. It can be observed that the MD detector outperforms ZF 
detector by about 8 dB at a BER of $10^{-5}$. DCM-OFDM
with MD detector outperforms QCM-OFDM with MD detector at lower to
moderate SNR values. However, QCM-OFDM with ZF detector outperforms
DCM-OFDM with ZF detector at all SNRs. The ZF detector performs well for
QCM-OFDM due to the availability of the additional degrees of freedom in
QCM. 

\section{Proposed SM-DCM scheme}
\label{sec5}
The DCM scheme proposed in the previous section conveys the magnitude 
and phase of a complex symbol through two LEDs. The achieved rate can 
be increased through the use of spatial modulation in DCM. The spatial 
modulation-DCM (SM-DCM) scheme proposed in this section is based on this 
idea. In the SM-DCM scheme, in addition to modulation bits, there are 
additional bits which are conveyed through spatial indexing. The DCM 
scheme restricts the number of LEDs to be two, whereas SM-DCM scheme 
allows the use of multiple pairs of LEDs. The proposed SM-DCM scheme 
uses two LEDs (one pair of LEDs) as one BLOCK to convey a complex symbol 
(just like in case of DCM). The number of BLOCKs (pairs of LEDs) is given 
by $N_p=\left \lfloor{\frac{N_t}{2}}\right\rfloor$. The selection of a 
BLOCK to use in a given channel use is done using index bits. The number 
of index bits for BLOCK selection is $\mbox{log}_2N_p$.

\subsection{SM-DCM transmitter}
\label{sec5a}
\begin{figure}
\centering
\includegraphics[width=3.35in, height=1.65in]{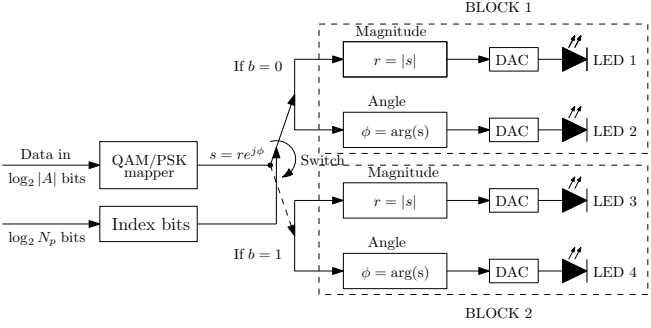}
\vspace{-2mm}
\caption{SM-DCM transmitter.}
\label{smdcmtx}
\vspace{-2mm}
\end{figure}

The block diagram of SM-DCM transmitter is shown in Fig. \ref{smdcmtx}.
In this scheme, we consider $N_t=4$. Therefore, the number of BLOCKs is, 
$N_p = 2$ (BLOCK 1 and BLOCK 2). After the data bits are mapped to a 
complex symbol (QAM/PSK), selection of the BLOCK to which this complex 
symbol is given as input is to be done. It is done by the index bits 
(in this case, it is one index bit, $b$) as follows.
\begin{equation}
\mbox{If} \hspace{2mm}
b = \left\{\begin{matrix}
0, &\mbox{then} &s \mbox{ goes to BLOCK 1} \\
1, &\mbox{then} &s \mbox{ goes to BLOCK 2}.
\end{matrix}\right.
\end{equation}
After the BLOCK selection is done, the process is same as that of the DCM 
scheme described in Sec. \ref{sec4}. If BLOCK 1 is selected, then LED1 
transmits the magnitude ($r$) and LED2 transmits the phase ($\phi$) of 
the complex symbol ($s$). Similarly, if BLOCK 2 is selected, then LED3 
transmits the magnitude ($r$) and LED4 transmits the phase ($\phi$) of 
the complex symbol ($s$). The definitions of $r$ and $\phi$ are same as 
in DCM.

{\em Example 3:} If the complex modulation symbol to be transmitted is
$s=3+\mbox{j}3$ and if the index bit is 0, then $r=3\sqrt{2}$, 
$\phi=\frac{\pi}{4}$ and BLOCK 1 will be selected. LED1 emits light of 
intensity $3\sqrt{2}$ and LED2 emits light of intensity $\frac{\pi}{4}$. 
LED3 and LED4 will be OFF. The corresponding SM-DCM transmit vector is
${\bf x}=[3\sqrt{2}\hspace{2mm}\frac{\pi}{4}\hspace{2mm} 0\hspace{2mm} 0]^T$.

{\em Example 4:} If the complex modulation symbol to be transmitted is
$s=3+\mbox{j}3$ and if the index bit is 1, then $r=3\sqrt{2}$, 
$\phi=\frac{\pi}{4}$ and BLOCK 2 will be selected. LED3 emits light of 
intensity $3\sqrt{2}$ and LED4 emits light of intensity $\frac{\pi}{4}$. 
LED1 and LED2 will be OFF. The corresponding SM-DCM transmit vector is
${\bf x}=[0\hspace{2mm} 0\hspace{2mm} 3\sqrt{2}\hspace{2mm} \frac{\pi}{4}]^T$.

\subsection{SM-DCM signal detection}
\label{sec5b}
\begin{figure}
\centering
\includegraphics[width=2.5in, height=1.25in]{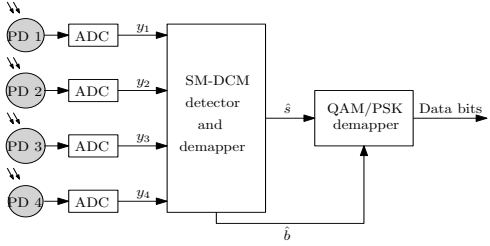}
\vspace{-2mm}
\caption{SM-DCM receiver.}
\label{smdcmrx}
\vspace{-2mm}
\end{figure}

The block diagram of SM-DCM receiver with $N_r=4$ PDs is shown in the 
Fig. \ref{smdcmrx}. Following the system model in Sec. \ref{sec2}, the 
$N_r\times 1$ received signal vector at the output of the PDs is given 
by (\ref{sysmodel}). Assuming perfect channel knowledge at the receiver, 
the ML estimate of the transmit vector ${\bf x}$ is obtained as
\begin{equation}
\hat{\bf x}_{ML} = \argmin_{{\bf x}\in {\mathbb S}_{SD}} \|{\bf y} - a{\bf Hx}\|^2,
\end{equation}
where ${\mathbb S}_{SD}$ denotes the SM-DCM signal set (consisting of all
possible ${\bf x}$ vectors). The detected vector $\hat{\bf x}_{ML}$ is
demapped to the corresponding complex symbol $\hat{s}_{ML}$, which is
then demapped to get the corresponding modulation bits. By looking at the 
non-zero indices of the detected vector $\hat{\bf x}_{ML}$, we can detect 
the index bits $\{\hat{b}\}$.

\textit{Achieved rate in SM-DCM}: 
The achieved rate in DCM is $\mbox{log}_2{|\mathbb A|}$ bpcu. It is 
increased by $\mbox{log}_2{N_p}$ due to index bits in SM-DCM. Thus, 
the achieved rate in SM-DCM is given by
\begin{equation}
\eta_{smdcm} =  \mbox{log}_2{|\mathbb A|} + \mbox{log}_2{N_p}\ \mbox{bpcu}.
\end{equation}

\subsection{Optimum placement of LEDs for SM-DCM}
\label{sec5c}
\begin{figure}
\center
\subfigure{\includegraphics[height=1.25in]{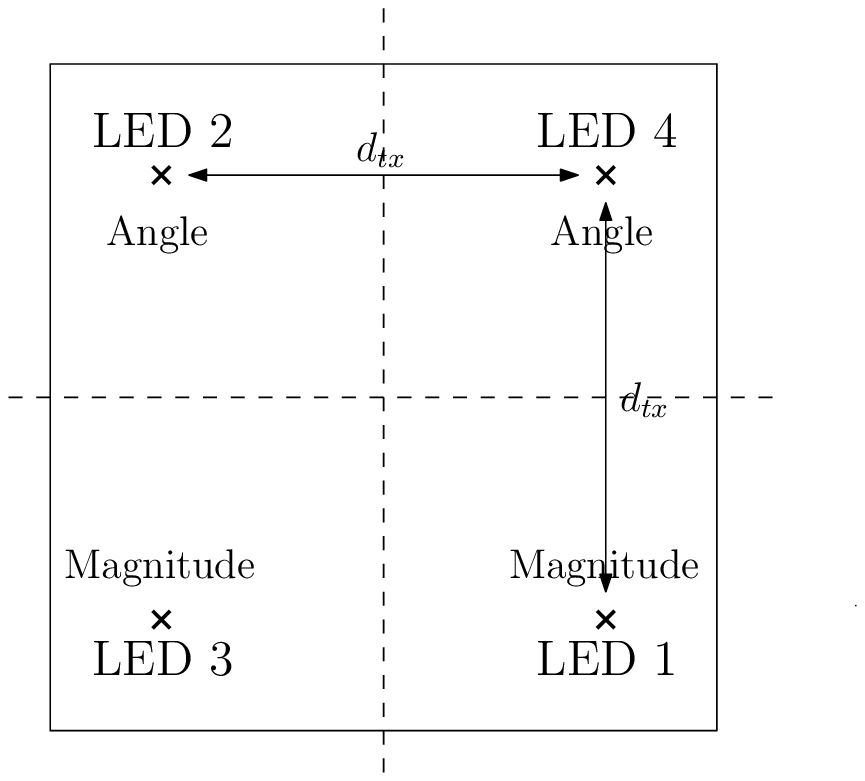}}
\hspace{3mm}
\subfigure{\includegraphics[height=1.25in]{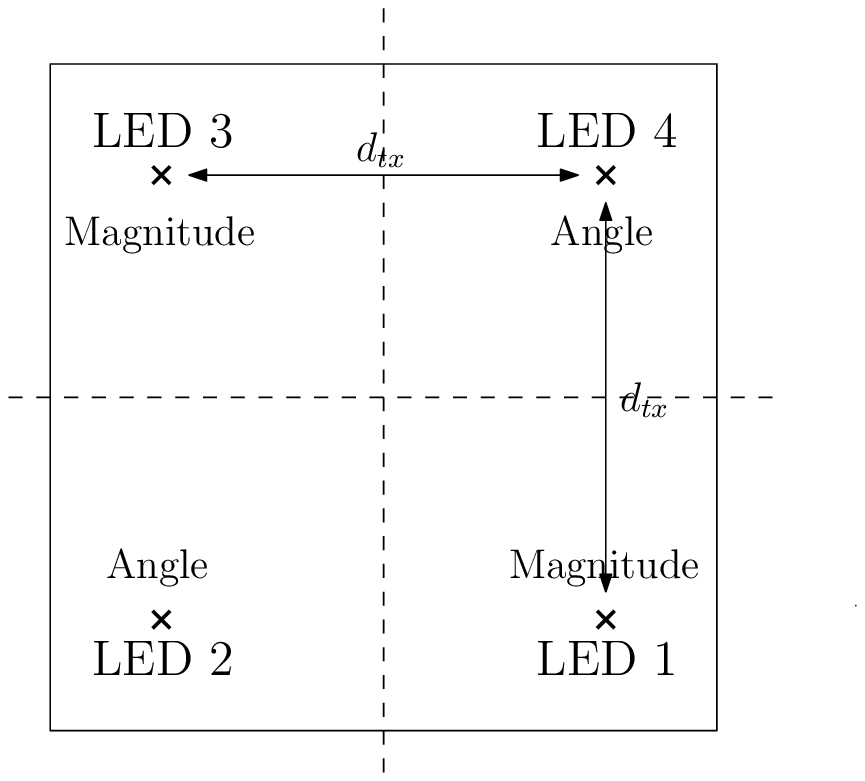}}

{\footnotesize \hspace{-2mm} (a) Placement P1 \hspace{20mm} (b) Placement P2}
\caption{Placement of LEDs for SM-DCM.}
\label{smdcmplcmt}
\end{figure}

Several LED placements are possible for SM-DCM. We use two metrics, 
namely, minimum Euclidean distance and average Euclidean distance 
between any two SM-DCM signal vectors $\vx_1$ and $\vx_2$ transmitted 
through $\mh$ (i.e., $d_{min,\textbf{H}}$ and $d_{avg,\textbf{H}}$), to 
decide the optimum placement of LEDs for SM-DCM. The definitions of 
$d_{min,\textbf{H}}$ and $d_{avg,\textbf{H}}$ are given below:
\begin{equation}
d_{min,\mh} \Define\argmin_{{\vx _1} ,{\vx _2}\in {\mathbb S}_{SD} } \ \|\mh({\vx _2}-{\vx _1})\|^2,
\label{dmin}
\end{equation}
\begin{equation}
d_{avg,\mh}\Define\frac{1}{{|{\mathbb S}_{SD}| \choose 2}}\sum_{{\vx _1} ,{\vx _2}\in {\mathbb S}_{SD} }\big \|\mh({\vx _2}-{\vx _1})\big \|^2 .
\label{davg}
\end{equation}
We choose the placement of LEDs such that $d_{min,\textbf{H}}$ and 
$d_{avg,\textbf{H}}$ in that placement are maximized. The placement of 
LEDs in Fig. \ref{smdcmplcmt} (a) and Fig. \ref{smdcmplcmt} (b) are 
considered as P1 and P2, respectively. We have computed values of 
$d_{min,\textbf{H}}$, $d_{avg,\textbf{H}}$ for placements P1, P2 and 
8-QAM, 32-QAM for the SM-DCM scheme. The computed values are listed in 
Table \ref{smdcmtable}. From this table, we observe that 
$d_{min,\textbf{H}}$ and $d_{avg,\textbf{H}}$ values are maximum for 
LED placement P2. So, we choose LED placement P2 (as in 
Fig. \ref{smdcmplcmt} (b)) in our simulation results for SM-DCM.

\begin{table}[h]
\centering
\begin{tabular}{|c|c|c|c|}
\hline
LED placement & Modulation & $d_{min,\textbf{H}}$ & $d_{avg,\textbf{H}}$ \\
\hline\hline
P1  & 8-QAM & 2.0420$\times10^{-14}$  & 5.0340$\times10^{-10}$ \\ \hline
P2 & 8-QAM & 8.5358$\times10^{-13}$ & 5.0460$\times10^{-10}$ \\ \hline
P1  & 32-QAM & 1.8470$\times10^{-14}$  & 6.2289$\times10^{-10}$ \\ \hline
P2 & 32-QAM & 9.2177$\times10^{-14}$ & 6.2510$\times10^{-10}$ \\ \hline
\end{tabular}
\vspace{4mm}
\caption{Values of $d_{min,\textbf{H}}$, $d_{avg,\textbf{H}}$ for different 
placements and $M$-QAM alphabets of SM-DCM.}
\label{smdcmtable}
\vspace{-6mm}
\end{table}

\subsection{BER performance of SM-DCM}
\label{sec5d}
\begin{figure}
\centering
\includegraphics[width=3.5 in, height=2.5in]{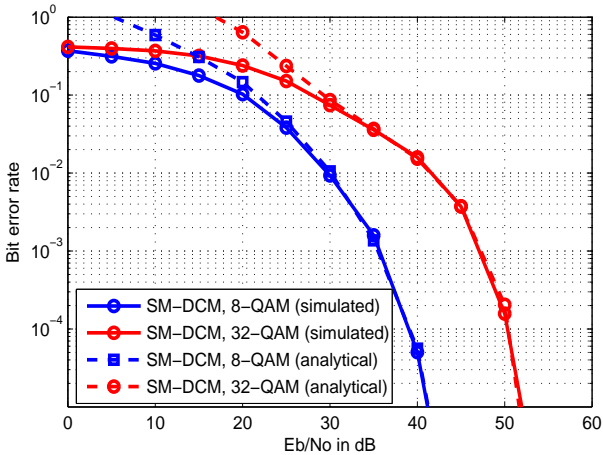}
\vspace{-4mm}
\caption{Comparison of analytical upper bound and simulated BER for 
SM-DCM with 8-QAM and 32-QAM.}
\label{fig19}
\vspace{-2mm}
\end{figure}

An upper bound on the BER of QCM for ML detection was obtained in Sec.
\ref{ub_qcm}. In a similar way, an upper bound on the BER of SM-DCM can
be obtained. Figure \ref{fig19} shows the upper bound and simulated BER 
plots for SM-DCM with ML detection at 8-QAM and 32-QAM. The system 
parameters used are the same as in Table \ref{tab1}. The placement of 
LEDs used for the simulations of SM-DCM is specified in Fig. 
\ref{smdcmplcmt} (b), and $d_{tx}=1$m. It can be seen that the upper 
bound is tight at moderate to high SNRs. In the next section 
(Sec. \ref{sec6}), we use this tight bound on BER to compute the 
achievable rate contours in SM-DCM.

\begin{figure}
\centering
\includegraphics[width=3.5 in, height=2.5in]{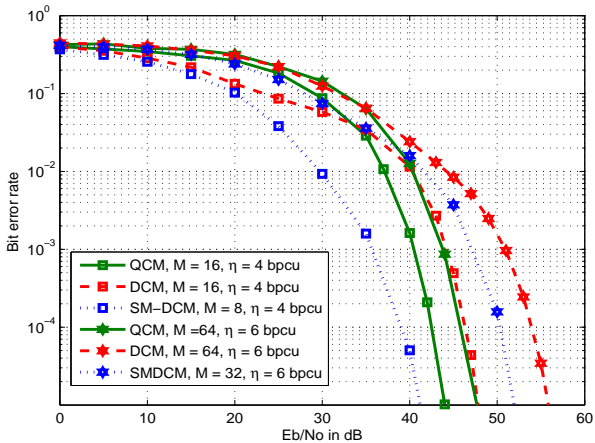}
\vspace{-4mm}
\caption{Comparison of BER performance of QCM, DCM, and SM-DCM for 
$\eta = 4, 6$ bpcu.}
\label{fig20}
\vspace{-2mm}
\end{figure}

Figure \ref{fig20} presents a comparison of the BER performance of QCM, 
DCM, and SM-DCM with ML detection for $\eta=4$ and 6 bpcu. For the DCM 
plots, LED1 and LED2 in Fig. \ref{smdcmplcmt} (b) are considered.
From Fig. \ref{fig20}, we observe that for $\eta=4$ bpcu, QCM and DCM 
require 16-QAM whereas SM-DCM requires 8-QAM only, which is why SM-DCM 
performs better by about 3 dB and 7 dB at $10^{-4}$ BER compared to 
QCM and DCM, respectively. Similarly, for $\eta=6$ bpcu, SM-DCM performs 
better than DCM by about 4 dB, since DCM requires 64-QAM whereas SM-DCM 
requires 32-QAM only. For $\eta=6$ bpcu, QCM performs better by about 
10 dB and 5 dB at $10^{-4}$ BER compared to DCM and SM-DCM, respectively. 
In case of QCM and SM-DCM for $\eta=6$ bpcu, SM-DCM performs better at 
low $E_b/N_0$ because of lower QAM size whereas QCM performs better 
at high $E_b/N_0$, because of the same reason as explained in Sec. 
\ref{sec4a}. That is, for large QAM sizes, the average relative distance 
between the transmit vectors in QCM is more compared to that in SM-DCM, 
i.e., if ($\vx_i, \vx_j$), $i\neq j$, is a pair of vectors from the 
transmit signal set, then $\mathbb{E}[\|\vx_i-\vx_j\|]$ is larger for 
the QCM signal set than for the SM-DCM signal set.

\section{Spatial distribution of SNR and achievable rate contours}
\label{sec6}
In addition to the BER vs $E_b/N_0$ performance plots presented in
the previous chapters, spatial characterization of the achieved
performance in the proposed QCM, DCM, and SM-DCM schemes is of interest. 
To address this need, in this section we study $i)$ the spatial
distribution of average received SNR ($\overline{\gamma}$ defined
in Sec. \ref{sec2}), $ii$) achievable rate contours for a given
target BER, and $iii)$ percentage area of the room covered for a
given rate and target BER in QCM, DCM, and SM-DCM. The system 
parameters used in the computation of the above quantities are as 
in Table \ref{tab1}. Additional system parameters are listed in Table
\ref{new_tab}. The LEDs placement and the signal mapping to LEDs for 
QCM and DCM are as in the Fig. \ref{fig5}. The LEDs placement and the 
signal mapping to LEDs for SM-DCM are as in the Fig. \ref{smdcmplcmt}. 
The output power of each LED is 1 Watt. We compute the average received 
SNRs at various spatial points on the plane of the receiver at a spatial
resolution of 2.5 cm. To compute the noise power at the receiver, we use 
the following expression given in \cite{new2} and use the parameter values 
used therein (which are summarized in Table \ref{new_tab}): 
\begin{equation}
\sigma^2=2qa(P_r+I_a/a)I_2/T+B_a\rho^2.
\label{npower}
\end{equation}

\begin{table}[t]
\centering
\begin{tabular}{|c||l|c|}
\hline
Parameter & Description & Value\\
\hline\hline
$q$ & Charge of an electron & $1.602\times 10^{-19}$ C \\ \hline
$I_a$ & Ambient light photo current & 5.84 mA \\ \hline
$I_2$ & Noise bandwidth factor & 0.562 \\ \hline
$T$ & Signaling interval & 0.05 $\mu$sec \\ \hline
$B_a$ & Photo diode amplifier bandwidth & 50 MHz \\ \hline
$\rho$ & Photo diode amplifier noise density & 5pA/$\sqrt{\textrm{Hz}}$\\\hline
\end{tabular}
\vspace{2mm}
\caption{\label{new_tab} System parameters for computation of noise power.}
\vspace{-6mm}
\end{table}

We consider a target BER of $10^{-5}$. Note that Figs. \ref{fig17}, 
\ref{fig18}, and \ref{fig19} demonstrated the tightness of the BER upper 
bounds obtained in Secs. \ref{sec3c}, \ref{sec4a}, and \ref{sec5d} for 
QCM, DCM, and SM-DCM, respectively. Indeed, the upper bounds and 
simulation results almost match for BERs below $10^{-3}$. Therefore, 
these bounds can be used to accurately map the spatial distribution of 
the SNRs to achievable rate contours for the considered target BER of 
$10^{-5}$. This is done as follows. Using the average received SNR at 
a given spatial position of the receiver and the BER vs SNR relation 
given by the BER upper bound expression, determine the maximum QAM size 
(among 2-, 4-, 8-, 16-, 32-, 64-QAM) that meets the $10^{-5}$ BER target. 
This determination is made for all spatial positions of the receiver at 
a spatial resolution of 2.5cm. The resulting spatial map of the maximum 
QAM size possible gives the achievable rate in bpcu at various spatial 
positions of the receiver.

\begin{figure*}
\center
\subfigure{\includegraphics[width=2.25in,height=5.0in]{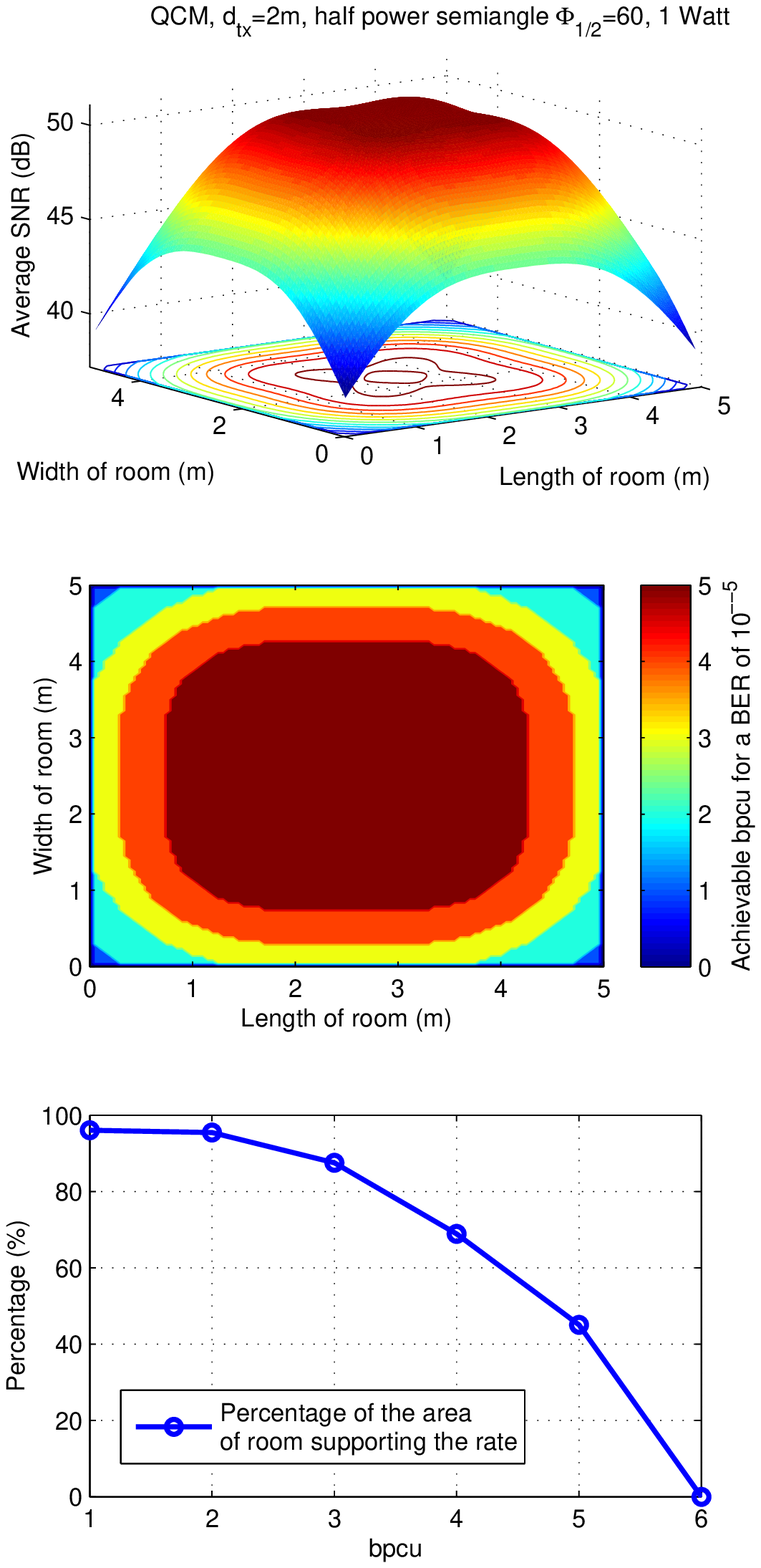}}
\hspace{2mm}
\subfigure{\includegraphics[width=2.25in,height=5.0in]{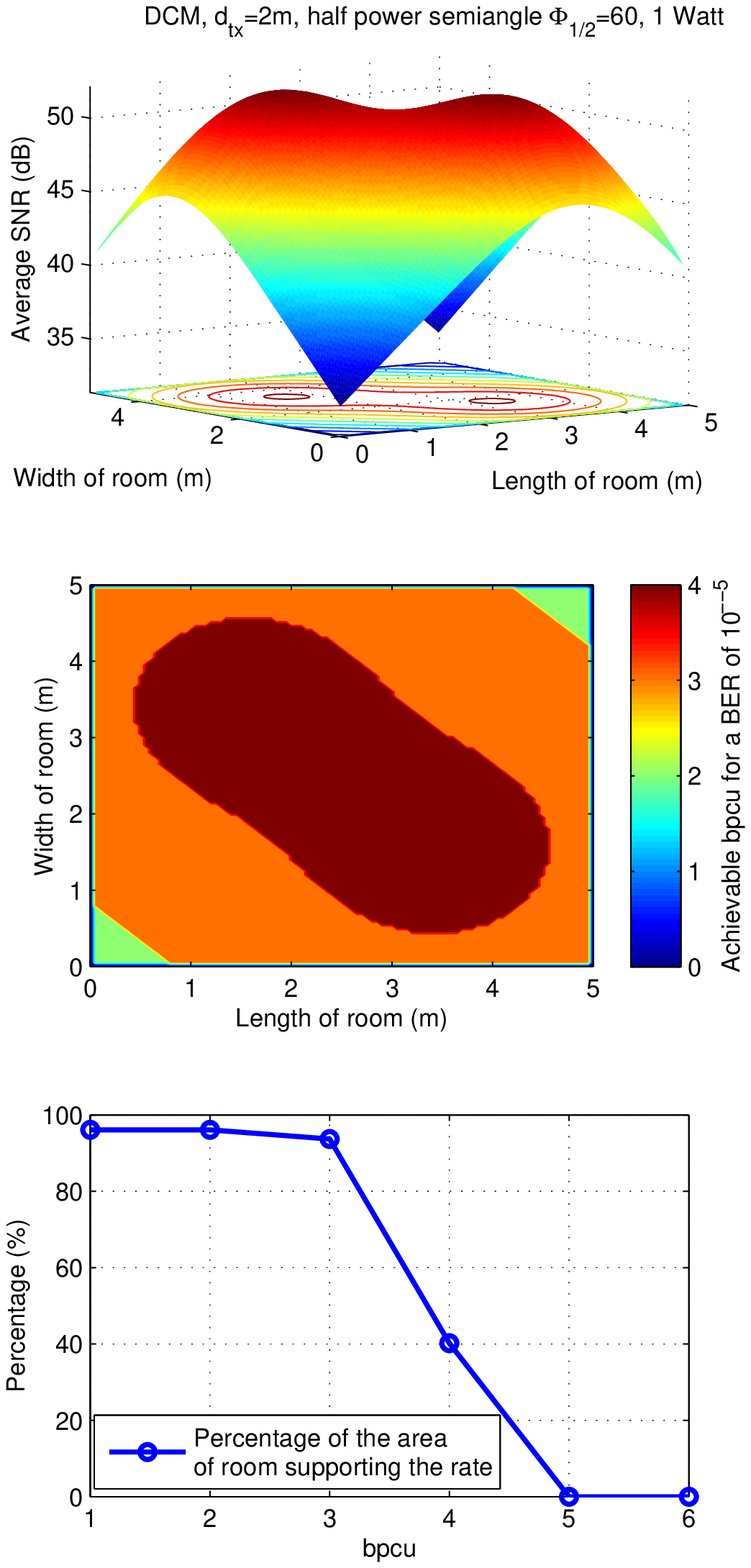}}
\hspace{2mm}
\subfigure{\includegraphics[width=2.25in,height=5.0in]{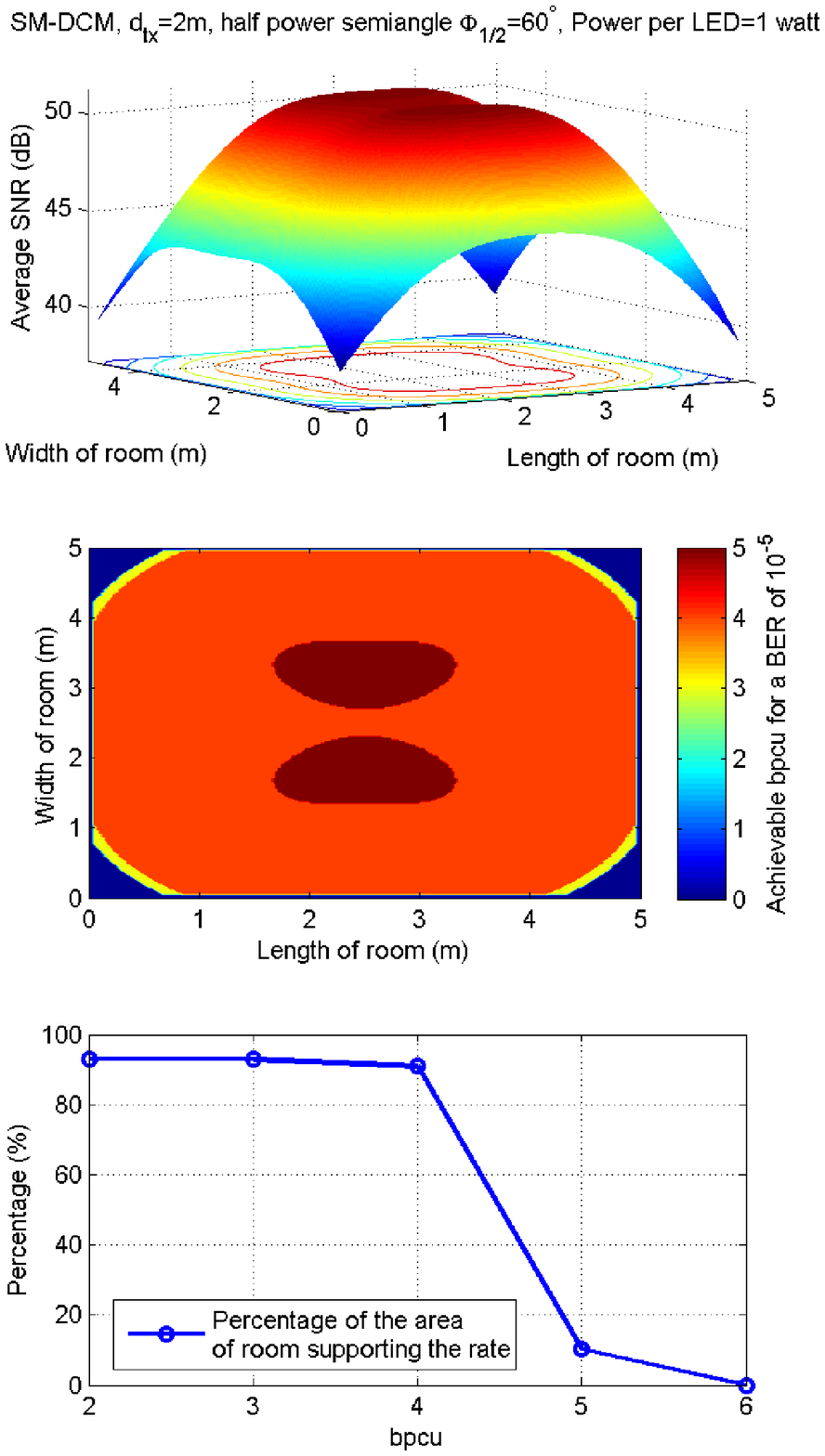}}

\vspace{-7mm} 
{\footnotesize (a) QCM \hspace{50mm} (b) DCM \hspace{50mm} (c) SM-DCM} 
\caption{Plots of spatial distribution of average received SNR, achievable
rate contours for $10^{-5}$ BER, and percentage area of the room covered vs
achieved rate for (a) QCM, (b) DCM, and (c) SM-DCM.}
\label{fig_a}
\vspace{-2mm} 
\end{figure*}

{\em Results and discussions:}
We computed the spatial performance measures discussed above for
QCM, DCM, and SM-DCM with $d_{tx}=2$m and $\Phi_{\frac{1}{2}}=60^\circ$.
Figures \ref{fig_a}(a),(b), and (c) show these performance plots for QCM, 
DCM, and SM-DCM, respectively. It can be observed that the maximum
rate achieved by QCM and SM-DCM while meeting the $10^{-5}$ BER target is 
5 bpcu (i.e., maximum supported QAM size is 32-QAM and 16-QAM for QCM and 
SM-DCM, respectively) and the maximum rate achieved by DCM is 4 bpcu (i.e., 
maximum supported QAM size is 16-QAM). This is due to the observation we 
made in Fig. \ref{dcmber1} and \ref{fig20}, where we saw that QCM had a 
larger average relative distance between the transmit vectors compared to 
DCM and SM-DCM for large QAM sizes and this resulted in a favorable 
performance for QCM over DCM and SM-DCM. This is found to result in QCM 
achieving a larger percentage area of the room covered by 4 bpcu (covering 
70\% area) and 5 bpcu (covering 45\% area) rates than DCM. Similarly, QCM 
achieves a larger percentage area of the room covered by 5 bpcu (covering 
70\% area) than SM-DCM. DCM shows a performance advantage over QCM for 
8-QAM; this can be seen by observing that DCM supports 8-QAM in more 
than 90\% of the room while QCM covers a lesser area for 8-QAM. Similarly, 
SM-DCM shows a performance advantage over QCM and DCM for $\eta=3, 4$ 
bpcu and $\eta=4$ bpcu, respectively. This can be seen by observing that 
SM-DCM covers more than 90\% of the room while QCM and DCM covers a 
lesser area for $\eta=3, 4$ bpcu and $\eta=4$ bpcu, respectively.

\section{Conclusions}
\label{sec7}
We proposed three simple and novel complex modulation schemes that 
avoided the Hermitian symmetry operation to generate LED compatible 
positive real signals encountered in VLC. This was achieved through 
the exploitation of the spatial dimension for the purpose of complex 
symbol modulation. In the proposed QCM scheme, four LEDs were used to 
convey the real and imaginary parts of a complex symbol and their sign 
information. While intensity modulation of LEDs was employed to convey 
the magnitudes of the real and imaginary parts, spatial index modulation 
of LEDs was used to convey their sign information. The proposed DCM 
scheme, on the other hand, exploited the polar representation of complex 
symbols to use only two LEDs to convey the magnitude and phase information 
of a complex symbol. The proposed SM-DCM scheme exploited the use of 
spatial modulation in DCM. Analytical upper bounds and simulation results 
showed that the proposed QCM, DCM, and SM-DCM achieve good BER performance. 
Phase rotation of modulation symbols was shown to improve the BER 
performance in QCM. Zero-forcing and minimum distance detectors for QCM 
and DCM when used along with OFDM showed good performance for these 
QCM-OFDM and DCM-OFDM schemes. The analytical BER upper bounds were 
shown to be very tight at high SNRs, and this enabled us to easily 
compute and plot the achievable rate contours for a given target BER 
(e.g., $10^{-5}$ BER) in QCM, DCM, and SM-DCM.

\end{document}